\documentclass[11pt,a4paper]{article}
\usepackage{jcappub}
\usepackage[compat=1.1.0]{tikz-feynhand}
\usepackage{bm}
\usepackage{amsmath}
\usepackage{mathrsfs}
\usepackage[mathscr]{euscript}
\usepackage{feynmp}
\usepackage{comment}
\usepackage{natbib}
\usepackage{ulem}

\newcommand{\Mpl}{M_\mathrm{Pl}}
\newcommand{\fR}{f_{\rm R}}
\newcommand{\lambdaR}{\lambda_{\rm R}}
\newcommand{\mR}{m_{\mathrm R}}

\title{Multi-chaotic inflation with and without spectator field}

\author[1]{Yukiyoshi Morishita,}
\author[2]{Tomo Takahashi,}
\author[3,4]{and Shuichiro Yokoyama}
\affiliation[1]{Department of Physics, Nagoya University, Nagoya 464-8602, Japan}
\affiliation[2]{Department of Physics, Saga University, Saga 840-8502, Japan}
\affiliation[3]{Kobayashi-Maskawa Institute, Nagoya University, Aichi 464-8602, Japan}
\affiliation[4]{Kavli Institute for the Physics and Mathematics of the
  Universe (WPI), The University of Tokyo,
  Kashiwa 277-8583, Japan}
\emailAdd{morishita.yukiyoshi``at"nagoya-u.jp}
\emailAdd{tomot``at"cc.saga-u.ac.jp}
\emailAdd{shu``at"kmi.nagoya-u.ac.jp}

\abstract{
Motivated by the result of Planck+BICEP/Keck recently released, we investigate the consistency of the multi-field inflation models in terms of the spectral index $n_s$ and the tensor-to-scalar ratio $r$. In this study, we focus on double-inflaton models with and without a spectator field.
We find that inflaton with a quadratic potential can become viable
when three fields with a specific hierarchical mass spectrum are realized such that two fields act as inflatons and the other one is the spectator. We also discuss the conditions to avoid the fine-tuning, by careful study of how the prediction depends on the background trajectory in the inflaton-field space. 
}


\usepackage{amsfonts}

\begin{document}

\begin{flushright}
\end{flushright}

\maketitle


\section{Introduction}
\label{sec:intro}

The inflation paradigm which predicts the accelerated expansion in the early Universe is widely accepted as the successful scenario. It solves problems in the standard Big Bang theory, however, the concrete mechanism of the inflation is yet to be understood. In the inflationary framework, the inflaton field $\phi$ is considered to drive the accelerated expansion and the validity of inflationary models has been tested by observations such as cosmic microwave background, large scale structure, and so on.  The spectral index $n_s$ has been severely constrained by Planck as $n_s = 0.9649 \pm 0.0042$ (68 \% C.L.) \cite{Planck:2018jri} and, the bound on the tensor-to-scalar ratio $r$ is recently updated by the observation of BICEP/Keck 2018 as $r_{0.05}<0.036$ (95 \% C.L.) \cite{BICEP:2021xfz}. 

Some models, such as chaotic inflation whose potential is given by $V(\phi)\propto\phi^p$, are ruled out by this recent result. However, adding another scalar field could change the predictions of the model. Actually, in high energy theories such as string theory, scalar fields are ubiquitous and multiple fields may be involved in the inflationary dynamics. Therefore, even if a model is not consistent with observational constraints as single-field inflation,  we can consider it in the multi-field framework, in which the predictions for the observable quantities might change and may become consistent with observational constraints.

Indeed there have been many works on multi-field inflation models, for example, in the context of double inflation  \cite{Silk:1986vc,Langlois:1999dw,Polarski:1992dq,Polarski:1994bk,Peter:1994dx,Polarski:1995zn,Hodges:1989um,Yamaguchi:2001pw,Feng:2003zua}, assisted inflation  \cite{Liddle:1998jc,Malik:1998gy,Copeland:1999cs,Coley:1999mj,Kaloper:1999gm}, {\it N}-flation \cite{Dimopoulos:2005ac,Easther:2005zr,Kim:2006ys,Kim:2006te,Piao:2006nm,Battefeld:2007en,Ashoorioon:2009wa,Ashoorioon:2011ki,Price:2014ufa,Guo:2017vxp}, multi-natural inflation \cite{Kim:2004rp,Czerny:2014wza,Choi:2014rja,Higaki:2014pja,Ben-Dayan:2014zsa}, multi-field monodromy inflation \cite{Wenren:2014cga}, and so on.  In these models, every field can be relevant to the inflationary dynamics in some way, i.e., all fields may be regarded as the inflatons. On the other hand, one can also consider the  multi-field model where a scalar field, called the spectator field, exists.
The spectator field is irrelevant to the inflationary dynamics but affects primordial density fluctuations. Examples of models of this type include the curvaton \cite{Enqvist:2001zp,Lyth:2001nq,Moroi:2001ct}, modulated reheating scenario \cite{Dvali:2003em,Kofman:2003nx} and so on, and the predictions for observables such as the spectral index $n_s$ and the tensor-to-scalar ratio $r$ in this kind of models have been well investigated \cite{Langlois:2004nn,Moroi:2005kz,Moroi:2005np,Ichikawa:2008iq,Ichikawa:2008ne,Enqvist:2013paa,Fujita:2014iaa,Vennin:2015vfa}. Furthermore such multi-field models can also give characteristic predictions for primordial non-Gaussianity, which have been  discussed in many works (e.g.,  \cite{Vernizzi:2006ve,Huang:2009vk,Sasaki:2006kq,Seery:2005gb,Rigopoulos:2005ae,Battefeld:2006sz,Yokoyama:2007dw,Yokoyama:2007uu,Ichikawa:2008iq,Ichikawa:2008ne,Suyama:2010uj,Frazer:2011br,Kobayashi:2012ba,Enqvist:2013paa,Fujita:2014iaa,Vennin:2015vfa}).

In this paper, we discuss a range of multi-field inflation models, particularly in light of recent BICEP/Keck constraints. We consider three kinds of multi-field models: the double-inflaton model, the single-inflaton and the single-spectator model which we call the mixed-double model, and the double-inflaton and the single-spectator model,
which we call hierarchical-triple model. For the potential of the inflaton(s), we adopt the chaotic inflation type. The first and second models above are two-field ones, whose predictions for $n_s$ and $r$ have been discussed in several works \cite{Langlois:2004nn,Moroi:2005kz,Moroi:2005np,Ichikawa:2008iq,Ichikawa:2008ne,Enqvist:2013paa,Fujita:2014iaa,Wenren:2014cga}. In this paper, we revisit these models in the light of recent BICEP/Keck results. The last model is a three-field one, which might be related to and motivated by some concrete particle physics setup \cite{Ellis:2013iea}, and is the main topic of this paper. We argue that, even though the recent Planck+BICEP/Keck constraint on $n_s$ and $r$ is severe, we can easily 
construct the models consistent with the recent constraints.

The contents of this paper are as below. In Sec.~\ref{double}, we first introduce the generalized chaotic inflation. Then we focus on the double chaotic inflation, in which all scalar fields are regarded as inflatons. In Sec.~\ref{triple}, we consider a model with the inflaton(s) and the spectator field. This section includes the main topic of this study. In Sec.~\ref{conclusion}, we give the conclusion of this paper, and implications of our analysis to some particle physics models are also briefly discussed.

\section{Double monomial inflation}
\label{double}

As we have mentioned in the introduction, the current observational constraints on the scalar spectral index $n_s$ and the tensor-to-scalar ratio $r$ imply that a class of the single field inflation model with a monomial potential is now excluded, and thus we are faced with the need to extend the model if we adopt the monomial potential. Here, as a possible minimal extension, we consider the double-inflaton model where two inflatons have both monomial potentials. 

First, for comparison, let us briefly discuss the predictions in the standard single slow-roll inflation with the monomial potential: 
\begin{eqnarray}
V=\lambda\frac{\phi^p}{\Mpl^{p-4}} \quad (p > 0),
\label{single_V}
\end{eqnarray}
with $\lambda$ and $p$ being dimensionless parameters. $\Mpl$ is the reduced Planck mass. The number of $e$-folds can be given by
\begin{eqnarray}
N(t) \equiv \int^{t_{\rm{end}}}_{t}H dt\simeq \frac{1}{\Mpl^2}\int_{\phi_{\rm{end}}}^{\phi}\frac{V}{V_{,\phi}}d\phi\simeq \frac{\phi^2}{2p\Mpl^2},
\label{single_enumber}
\end{eqnarray}
where $V_{,\phi} \equiv d V(\phi) / d \phi$.
Here we use the slow-roll approximation and assume that $\phi (t) \gg \phi_{\rm end}$. Employing the $\delta N$ formalism \cite{Starobinsky:1985ibc,Salopek:1990jq,Sasaki:1995aw,Sasaki:1998ug,Lyth:2004gb}, 
we can evaluate the power spectrum 
of the curvature perturbations as 
\begin{eqnarray}
\mathcal{P}_{\zeta}(k)
=N_{,\phi_\ast}^2\left(\frac{H_*}{2\pi}\right)^2,
\label{single_pzeta}
\end{eqnarray}
where the subscript $\ast$ denotes the value evaluated at the time when the scale of interest exits the horizon
and $N_{,\phi_\ast} \equiv \partial N(t_\ast) / \partial \phi_\ast$ is the derivative of the number of $e$-folds with respect to
the inflaton field evaluated at the horizon crossing time (taken to be a flat hypersurface). 
By using the slow-roll formula for the number of $e$-folds, we can rewrite the power spectrum as
\begin{eqnarray}
\mathcal{P}_{\zeta}(k)
=\frac{2N_\ast}{p \Mpl^2 }\left(\frac{H_*}{2\pi} \right)^2 \,.
\label{single_pzeta_2}
\end{eqnarray}

On the other hand, the power spectrum of the tensor fluctuations generated during the inflation is given by
\begin{eqnarray}
\mathcal{P}_{\rm T}(k)=\frac{8}{\Mpl^2}\left(\frac{H_*}{2\pi}\right)^2.
\label{pg}
\end{eqnarray}
Using these quantities, we can define and evaluate the spectral index and the tensor-to-scalar ratio as 
\begin{eqnarray}
n_s-1&\equiv& \frac{d\ln{\mathcal{P}_{\zeta}}}{d\ln{k}}
= -\frac{d\ln{\mathcal{P}_{\zeta}}}{dN_\ast}
=-\frac{1}{N_\ast}-2\epsilon_\ast
=-\frac{p+2}{2N_\ast},
\label{single_ns} \\
r&\equiv& \frac{\mathcal{P}_{\rm T}}{\mathcal{P}_{\zeta}}=\frac{4p}{N_\ast},
\label{single_r}
\end{eqnarray}
where $\epsilon$ is a slow-roll parameter and it can be written as 
\begin{eqnarray}
\epsilon \equiv \frac{d \ln H}{dN}
\simeq \frac{\Mpl^2}{2} \left(\frac{V_{,\phi}}{V}\right)^2
= \frac{p^2}{2} \frac{\Mpl^2}{\phi^2}
=\frac{p}{4N}.
\label{eq:slowepsilon}
\end{eqnarray}
Thus, for the single monomial inflation,
we can find a useful relation between $n_s$ and $r$ as
\begin{eqnarray}
r = -\frac{8p}{p+2} (n_s - 1).
\end{eqnarray}
In particular, for the case with $p=2$ and $N_\ast = 55$,
we have $n_s = 1 - 2/N_\ast \sim 0.964$
and $r \sim 0.15$, and for the case with $p=2/5$ and $N_\ast = 55$, we have $r \sim 0.029$ and $n_s \sim 0.978$. Therefore, the class of the single monomial inflation model is already excluded by the observational constraint from the recent  Planck+BICEP/Keck~2018 result \cite{BICEP:2021xfz}.

\subsection{Slow-roll formulae for the double monomial inflation}
\label{double_chaotic}

As an extension of the single-field monomial (chaotic) inflation model, let us consider the class of the double-inflaton model
where two scalar fields with the separable-monomial potential behave both as the slow-roll inflatons. Here, the potential is simply assumed to be
\begin{eqnarray}
V(\phi,\chi) = U(\phi)+ W(\chi)
=\lambda_\phi\frac{\phi^{p_\phi}}{\Mpl^{p_\phi-4}}+\lambda_\chi\frac{\chi^{p_\chi}}{\Mpl^{p_\chi-4}},
\label{double_V}
\end{eqnarray}
where $\lambda_\phi$ and $\lambda_\chi$ are dimensionless parameters. 
For simplicity, the power of monomials, $p_\phi$ and $p_\chi$, are assumed to be
the same: $p_\phi = p_\chi = p$.
We will briefly comment on the case with
$p_\phi \neq p_\chi$ later.
For such a separable form of the potential,
under the slow-roll approximation
the number of $e$-folds can be approximately
obtained as (see, e.g.,~\cite{Starobinsky:1985ibc})
\begin{eqnarray}
N(t)=\frac{1}{\Mpl^2}\left[\int_{\phi_{\rm{end}}}^{\phi}\frac{U(\phi)}{U_{,\phi}}d\phi+\int_{\chi_{\rm{end}}}^{\chi}\frac{W(\chi)}{W_{,\chi}}d\chi \right]
\simeq \frac{1}{2p \Mpl^2}(\phi^2+\chi^2) \,.
\label{double_enumber}
\end{eqnarray}

The curvature perturbations can be given by the sum of the contributions from the two inflatons, namely  $\zeta=\zeta^{(\phi)}+\zeta^{(\chi)}$ where $\zeta^{(\varphi_i)}$ is the curvature perturbations generated by $\varphi_i (=\phi, \chi)$, and adopting the $\delta N$ formalism and using the expression of the $e$-folds~\eqref{double_enumber}, the power spectrum is simply given by 
\begin{eqnarray}
\mathcal{P}_\zeta
=\mathcal{P}_\zeta^{(\phi)}+\mathcal{P}_\zeta^{(\chi)}
=\sum_i \left(\frac{\partial N_\ast}{\partial \varphi_{i\ast}}\right)^2 \left(\frac{H_*}{2\pi}\right)^2
=\frac{2N_\ast}{p \Mpl^2}\left(\frac{H_*}{2\pi}\right)^2,
\label{double_pzeta}
\end{eqnarray}
where $\varphi_{i\ast}=\phi_\ast, \chi_\ast$.\footnote{
Note that it can be easily verified that this expression holds in general even if there are more than two inflatons.
}
Since this expression is the same as in the single-field case and also the power spectrum of the tensor fluctuations is not modified, the slow-roll expressions for $n_s$ and $r$ in the double monomial case are virtually the same as those for the single-field case:
\begin{eqnarray}
n_s-1
&=& - \frac{1}{N_\ast} - 2\epsilon_\ast
\label{double_ns}, \\
r&=&\frac{4p}{N_\ast}.
\label{double_r} 
\end{eqnarray}
However, the slow-roll parameter, $\epsilon$,
would be modified from that in the single-field case and it depends on the trajectory in
the field space and it is given by
\begin{eqnarray}
\epsilon \simeq \frac{\Mpl^2}{2}
\sum_i \left(\frac{\partial V / \partial \varphi_i}{ V(\phi, \chi)}\right)^2
&=& \frac{\Mpl^2}{2} \frac{U_{,\phi}^2 + W_{,\chi}^2}{V^2} \cr\cr
&=& \frac{p^2}{2}\frac{\Mpl^2}{\phi^2}
 \frac{1+\left(\frac{\lambda_\chi}{\lambda_\phi}\right)^2 \left(\frac{\chi}{\phi}\right)^{2p-2}}{\left[1+\left(\frac{\lambda_\chi}{\lambda_\phi}\right) \left(\frac{\chi}{\phi}\right)^p\right]^2} \cr\cr
 &\simeq&
\frac{p}{4N}\left[1+\left(\frac{\chi}{\phi}\right)^2\right] \frac{1+\left(\frac{\lambda_\chi}{\lambda_\phi}\right)^2 \left(\frac{\chi}{\phi}\right)^{2p-2}}{\left[1+\left(\frac{\lambda_\chi}{\lambda_\phi}\right) \left(\frac{\chi}{\phi}\right)^p\right]^2},
\label{double_epsilon}
\end{eqnarray}
where we have used Eq.~\eqref{double_enumber}, and the modification from the single-field case can be characterized by the ratio between the constant (coupling) parameters, $\lambda_\phi$ and $\lambda_\chi$,
and also that between the field values $\phi$ and $\chi$.
Thus, let us introduce the parameters
$\lambdaR \equiv \lambda_\chi / \lambda_\phi$ and $\fR \equiv \chi_\ast / \phi_\ast$. By using these parameters, to see clearly the effect of two inflatons, we rewrite $\epsilon$ at the time of the horizon crossing
as
\begin{eqnarray}
\epsilon_\ast
=\frac{p}{4N_\ast}\left[1+\frac{\fR^2(1-\lambdaR\fR^{p-2})^2}{(1+\lambdaR\fR^p)^2}\right].
\label{double_epsilon2}
\end{eqnarray}
This equation shows that for the same $p$
and $N_\ast$
the slow-roll parameter $\epsilon_\ast$
in the double monomial inflation is always larger than that in the single-field case, and from the expressions of $n_s$ and $r$ given in Eqs.~\eqref{double_ns} and \eqref{double_r} respectively, we can find that, even for the same $p$ and $N_\ast$,  in the double monomial case $n_s$ can be smaller but $r$ is completely the same as that in the single-field case. 
As can be easily found in the above expression, for $p=2$ case
if $\lambdaR = 1$ the slow-roll parameter
$\epsilon$ is independent of the choice of $\fR$, that is, the choice of the background trajectory in the field space, and
it is completely the same as in the single-field case.
This is actually due to the exact $O(2)$ symmetry for the system of two scalar fields. 

Let us comment on the $n_{\rm T}-r$ consistency relation in the double monomial inflation.
As we have mentioned, 
the tensor fluctuations in both cases
are completely the same
and the spectral index of the tensor fluctuations is given as $n_{\rm T} = - 2 \epsilon_\ast$.
However, as shown above
the relation between the $\epsilon_\ast$
and $N_\ast$ would be modified in the double-inflaton case. This means that the consistency relation between $n_T$ and $r$ should be changed as
\begin{eqnarray}
\left\{ \,
\begin{aligned}
&r=- 8 n_{\rm T} \quad (\rm{single}) \,, \\
&r =- 8 n_{\rm T} \left[1+\frac{\fR^2(1-\lambdaR\fR^{p-2})^2}{(1+\lambdaR\fR^p)^2}\right]^{-1} \leq - 8 n_{\rm T} \quad (\rm{double}) \,.
\end{aligned}
\right.
\label{double_nt_r}
\end{eqnarray}

\subsection{Validity of the slow-roll formulae}
\label{sec:valid}

Here, we briefly discuss the validity of the above simple slow-roll formulae, by calculating the evolution of the adiabatic curvature perturbations numerically
based on the $\delta N$ formalism. 
In the $\delta N$ formalism \cite{Starobinsky:1985ibc,Salopek:1990jq,Sasaki:1995aw,Sasaki:1998ug,Lyth:2004gb},
when we numerically calculate $\mathcal{P}_\zeta$ at the leading order, we need to evaluate the difference of the number of $e$-folds measured from the initial time (taken to be the flat slicing) to the final time for the change of the field values at the initial time. As commonly assumed, we have taken the initial time to be the horizon crossing, $t_\ast$, and the final time to be the end of the inflation, and evaluated the difference of the number of $e$-folds by shifting the initial values of the inflatons
and
solving the two different background evolution equations  
as
\begin{eqnarray}
\delta N = \frac{\partial N_\ast}{\partial \phi_\ast} \delta \phi_\ast
+ \frac{\partial N_\ast}{\partial \chi_\ast} \delta \chi_\ast,
\end{eqnarray}
with
\begin{eqnarray}
\frac{\partial N_\ast}{\partial \phi_\ast}=\frac{N(\phi_\ast+\delta \phi_\ast)-N(\phi_\ast-\delta \phi_\ast)}{2\delta \phi_\ast}, \quad
\frac{\partial N_\ast}{\partial \chi_\ast}=\frac{N(\chi_\ast +\delta \chi_\ast)-N(\chi_\ast-\delta \chi_\ast)}{2\delta \chi_\ast},
\label{triple_Nderivative} 
\end{eqnarray}
where we choose the values of $\delta \phi_\ast$ and $\delta \chi_\ast$ to get $\delta N=\mathcal{O}(10^{-1})$.

For simplicity, here we take the quadratic potential, that is, $p=2$, to demonstrate the validity of the slow-roll formulae for the spectral index and the tensor-to-scalar ratio.\footnote{
Note that 
to perform the same calculation
for the case called monodromy inflation~\cite{Silverstein:2008sg,Wenren:2014cga} (e.g., $p=2/3, 2/5$), we need to specify the form of the potential around the minimum to avoid the singularity which appears in the simple potential \eqref{double_V}.}
\begin{figure}[htbp]
\begin{tabular}{c}
\begin{minipage}{0.45\hsize}\includegraphics[scale=0.7]{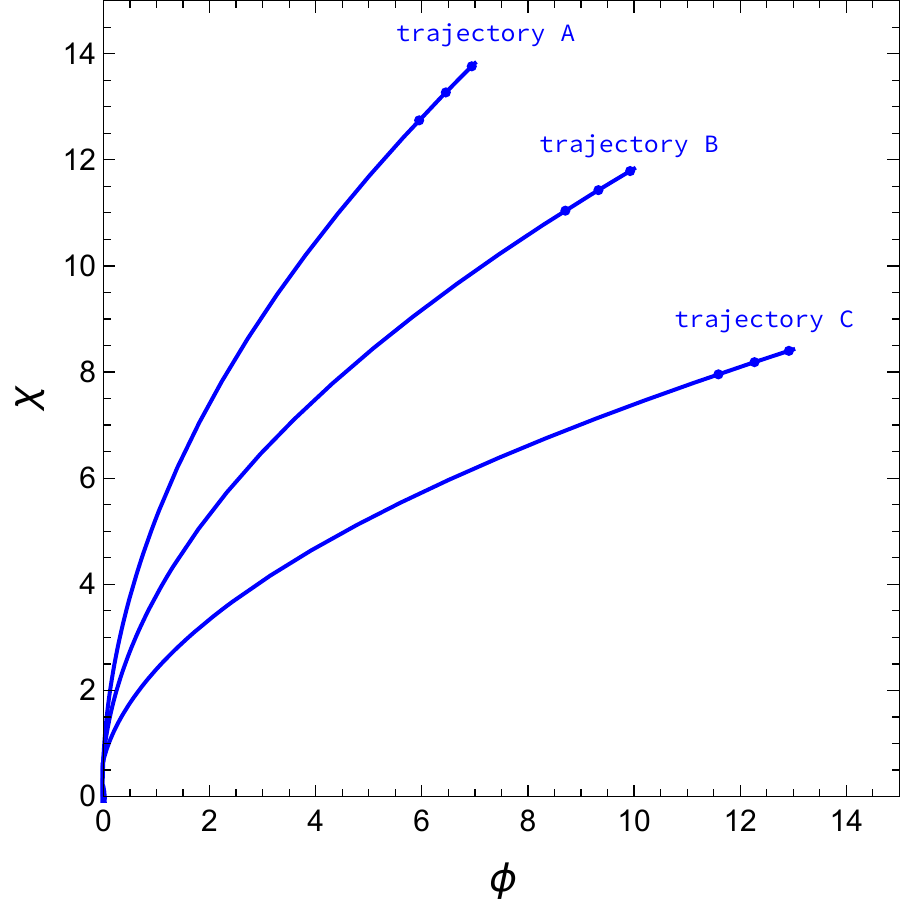}\end{minipage}
\begin{minipage}{0.5\hsize}\includegraphics[scale=0.7]{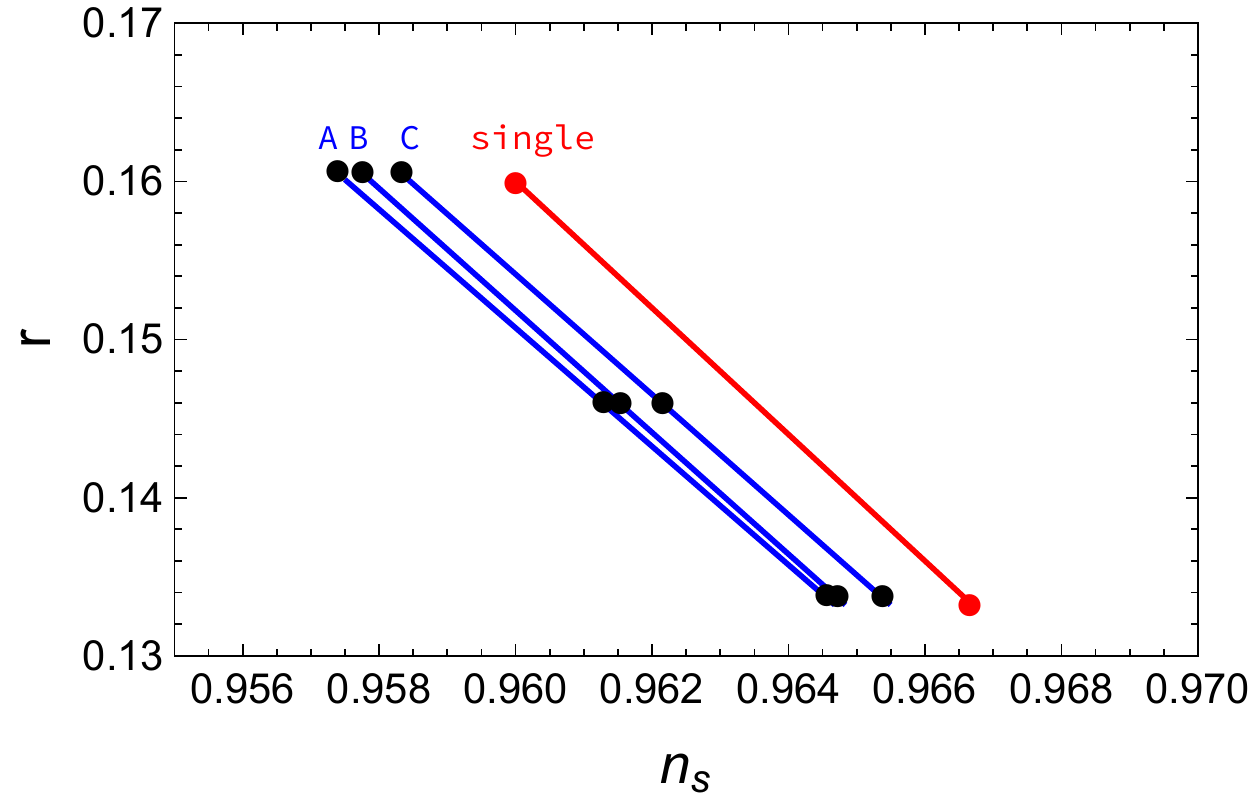}\end{minipage}
\end{tabular}
\caption{The left panel shows three trajectories, A, B, and C, in the field space. Here we fix $\lambdaR = 1/2$, which corresponds to the case where $\phi$-field is heavier than $\chi$-field. The points on each trajectory respectively correspond to $N_\ast =50, 55$, and $60$ from bottom to top.
The right panel shows the result of $n_s$ and $r$. The red line is corresponding to the prediction for the single case with $p=2$ and the top and bottom ends are respectively for $N_\ast = 50$ and $60$. The blue lines (A, B, and C) are drawn by using the slow-roll formulae \eqref{double_ns} and \eqref{double_r} with \eqref{double_epsilon}, and in these expressions we take the field values on the background trajectories from $N_\ast = 50$ to $60$. The black points are numerical results at the corresponding points on the trajectories shown in the left panel.
}
\label{fig_double_p2nsr}
\end{figure}
As shown in the left panel of Fig.~\ref{fig_double_p2nsr},
we take three background trajectories, A, B, and C, and the points on each trajectory respectively
correspond to the $N_\ast = 50$, $55$, and $60$
from bottom to top. 
Here, we fix $\lambdaR = 1/2$, that is,
$\phi$-field is heavier than $\chi$-field.
At the each point, we numerically calculate the above $\partial N_\ast / \partial \varphi_{i,\ast}$ and 
estimate the spectral index and the tensor-to-scalar ratio
which are shown in the right panel
as black points.
In the right panel, with the blue lines,
we also show the analytic result obtained
by the slow-roll formulae \eqref{double_ns} and \eqref{double_r} with \eqref{double_epsilon}.
To estimate the spectral index $n_s$,
we need to specify not only $N_\ast$ but also
the corresponding field values in the double-inflaton case, and then we take the field values on each background trajectory for corresponding $N_\ast$.
For comparison, we also show the result in the standard single quadratic case by the red line. From this figure, one can easily find that
the simple slow-roll formulae work well.
The difference between the numerical results
and the analytic formulae
is as small as the order of the slow-roll parameters, which gives the error in $n_s$ and $r$ at the second order in slow-roll and would be negligible. Furthermore, as mentioned above, in the double monomial inflation, depending on the trajectories in the field space, even for the same $N_\ast$, the spectral index is shifted to a smaller value compared with the single-field case. 
The larger the field value of the heavier field is, the more the double monomial case approaches to the single-field one, which means that
the heavier field mostly dominates the total energy density in that case.
On the other hand, the tensor-to-scalar ratio
is independent of the choice of the background trajectory and it is the same as in the single-field case.

In the use of the slow-roll formulae \eqref{double_ns} and \eqref{double_r}, we should be careful that each field value has a lower limit because we assume that the two scalar fields play the role of the inflaton and satisfy the slow-roll condition during inflation, that is,
\begin{eqnarray}
\epsilon_\phi \equiv \frac{\Mpl^2}{2}
\left( \frac{U_{,\phi}}{U} \right)^2 < 1 \quad {\rm and} \quad 
\epsilon_\chi \equiv \frac{\Mpl^2}{2}
\left( \frac{W_{,\chi}}{W} \right)^2 < 1 \,,
\end{eqnarray}
from which one finds that 
\begin{eqnarray}
\frac{1}{2}p^2\Mpl^2<\phi^2, \chi^2.
\label{double_baundary condition} 
\end{eqnarray}
Here we should also mention the absolute value of the constant (coupling) parameters. In the above formulae, not the absolute value but the ratio of $\lambda_\phi$ and $\lambda_\chi$, denoted by $\lambdaR$, affects $n_s$ and $r$. However, when the observed normalization of the amplitude for the scalar curvature perturbation, $\mathcal{P}_{\zeta, {\rm obs}}=2.2\times10^{-9}$,
is taken into account, we can specify the suitable absolute value for the constant parameter as
\begin{eqnarray}
\lambda_\phi=\frac{6p\pi^2\Mpl^2}{N_\ast}(2 p N_\ast\Mpl^2)^{-p/2}\frac{(1+\fR)^{p/2}}{1+\lambdaR \fR^p}\times\mathcal{P}_{\zeta, {\rm obs}} \,.
\label{double_lambda1}
\end{eqnarray}

\subsection{Double monomial inflation in light of the recent BICEP/Keck result}
\label{sec:lightBK}

Here we compare the predictions of the double monomial inflation with the recent result reported by BICEP/Keck collaboration~\cite{BICEP:2021xfz},
which gives a tight constraint on the
tensor-to-scalar ratio as
$r_{0.05} < 0.036$ at $95\%$ C.L.
Combining this result with the Planck 2018 result, the single-field monomial inflation is excluded. In particular, the monodromy type with
$p=2/3$ or $2/5$ predicts the consistent
tensor-to-scalar ratio but slightly larger values for the spectral index than the current bound. 

As shown in the previous section, however, by considering two scalar fields with the monomial potential as inflatons,
the prediction for the spectral index can be shifted to the left in the $n_s$-$r$ plane compared with the single case,
while the tensor-to-scalar ratio does not change.
\begin{figure}[htbp]
\begin{tabular}{c}
\begin{minipage}{\hsize}\includegraphics[scale=1]{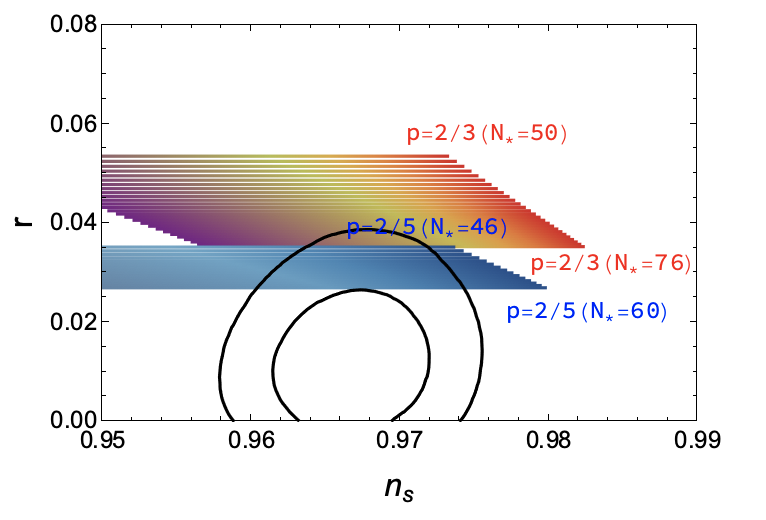}\end{minipage}
\end{tabular}
\caption{
Predictions of the double monomial inflation with $p=2/3$
(lines changing color from red to purple)
and $p=2/5$ (lines changing color from blue to light blue). 
$\lambdaR$ is fixed to be 1.
For the case of $p=2/3$, the number of $e$-folds $N_*$ varies from 50 to 76. On the other hand, for $p=2/5$, the range of $N_*$ is 46 to 60. The color of each line varies corresponding to the change of $\fR$. 
For each line, the right end corresponds to the single case. From Eq.~\eqref{double_epsilon2},
$\fR = 1$ is equivalent to the single case
in the prediction for $n_s$
and then in this plot, we change $\fR$ from $1$ to the maximum value that satisfies the slow-roll condition
given by Eq.~\eqref{double_baundary condition}.
The black solid lines correspond to the $1\sigma$
and $2\sigma$ constraints from Planck+BICEP/Keck~2018 \cite{BICEP:2021xfz}.
}
\label{fig_double_nsr}
\end{figure}
Thus, as shown in Fig.~\ref{fig_double_nsr}, we can construct an inflationary model consistent with the BICEP/Keck result with the double monodromy inflation~\cite{Wenren:2014cga}. 
In this plot, for each case, the right end corresponds to the prediction in the single case.
Even in the double monomial model,
from Eq.~\eqref{double_epsilon2},
it is found that by taking $\fR = 1$ with $\lambdaR = 1$ the prediction of $n_s$ is the same as in the single case. From the symmetry, the minimum value for $\fR$ can be regarded as 1, and hence we change $\fR$ just in the range larger than $1$ to the maximum value that is given as
\begin{eqnarray}
f_{\rm R, max} = \sqrt{\frac{4N_\ast}{p}-1}~,
\end{eqnarray}
obtained from the expression for the number of $e$-folds~\eqref{double_enumber} and the slow-roll condition \eqref{double_baundary condition}.
As we mentioned above, the constraint on the tensor-to-scalar ratio from the recent Planck+BICEP/Keck 2018 is $r_{0.05} < 0.036$ at $95\%$ C.L., from which one can read off the critical number of $e$-folds consistent with the upper bound on $r$ from Fig.~\ref{fig_double_nsr}. We need a large value as $N_*=76$ for $p=2/3$, and on the other hand, for the case of $p=2/5$, the range of $N_*$ consistent with the bound is $N_* >46$, which includes the reasonable value of $N_*=50 - 60$.

\begin{figure}[htbp]
\begin{tabular}{c}
\begin{minipage}{0.5\hsize}\includegraphics[scale=0.6]{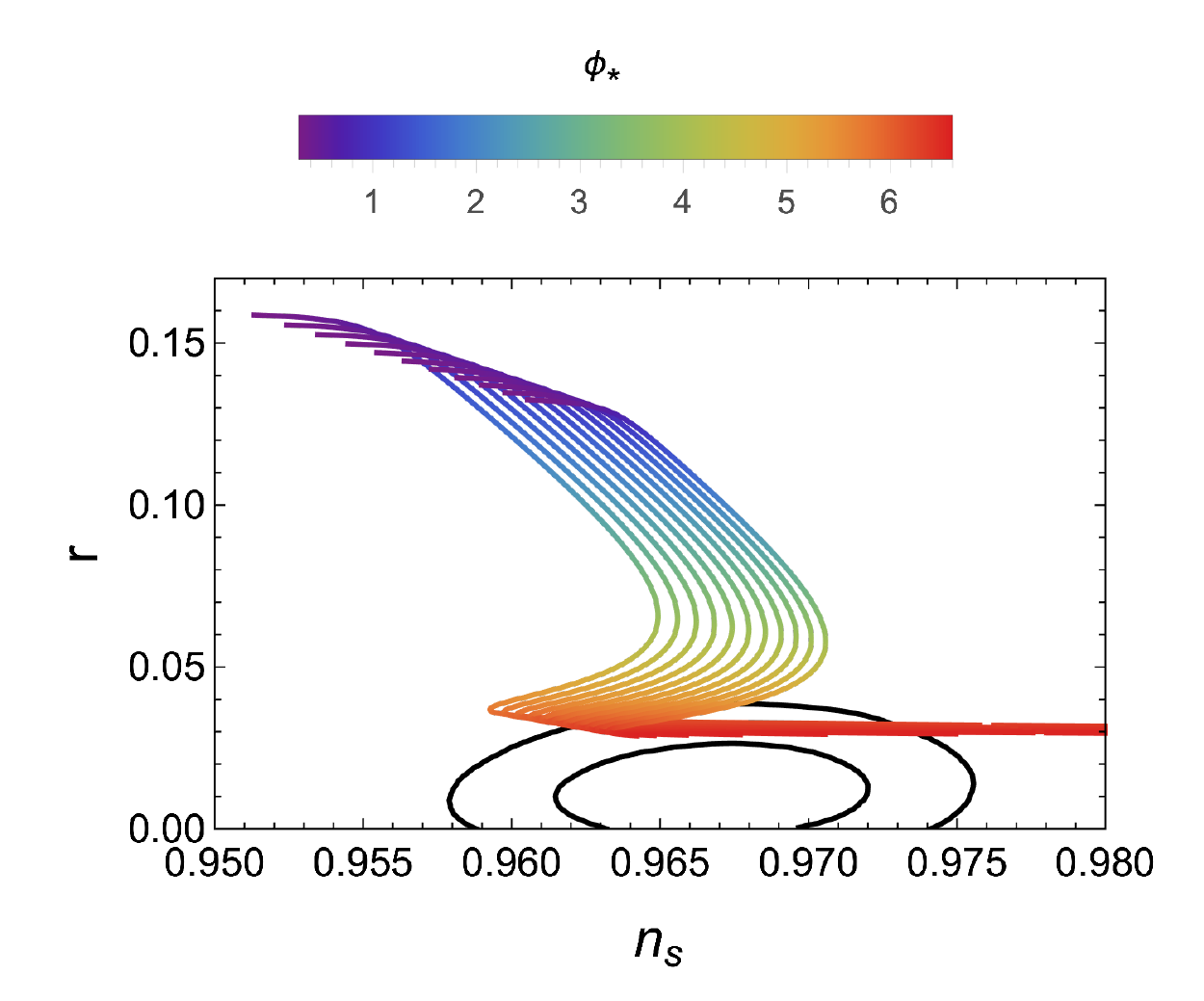}\end{minipage}
\begin{minipage}{0.5\hsize}\includegraphics[scale=0.5]{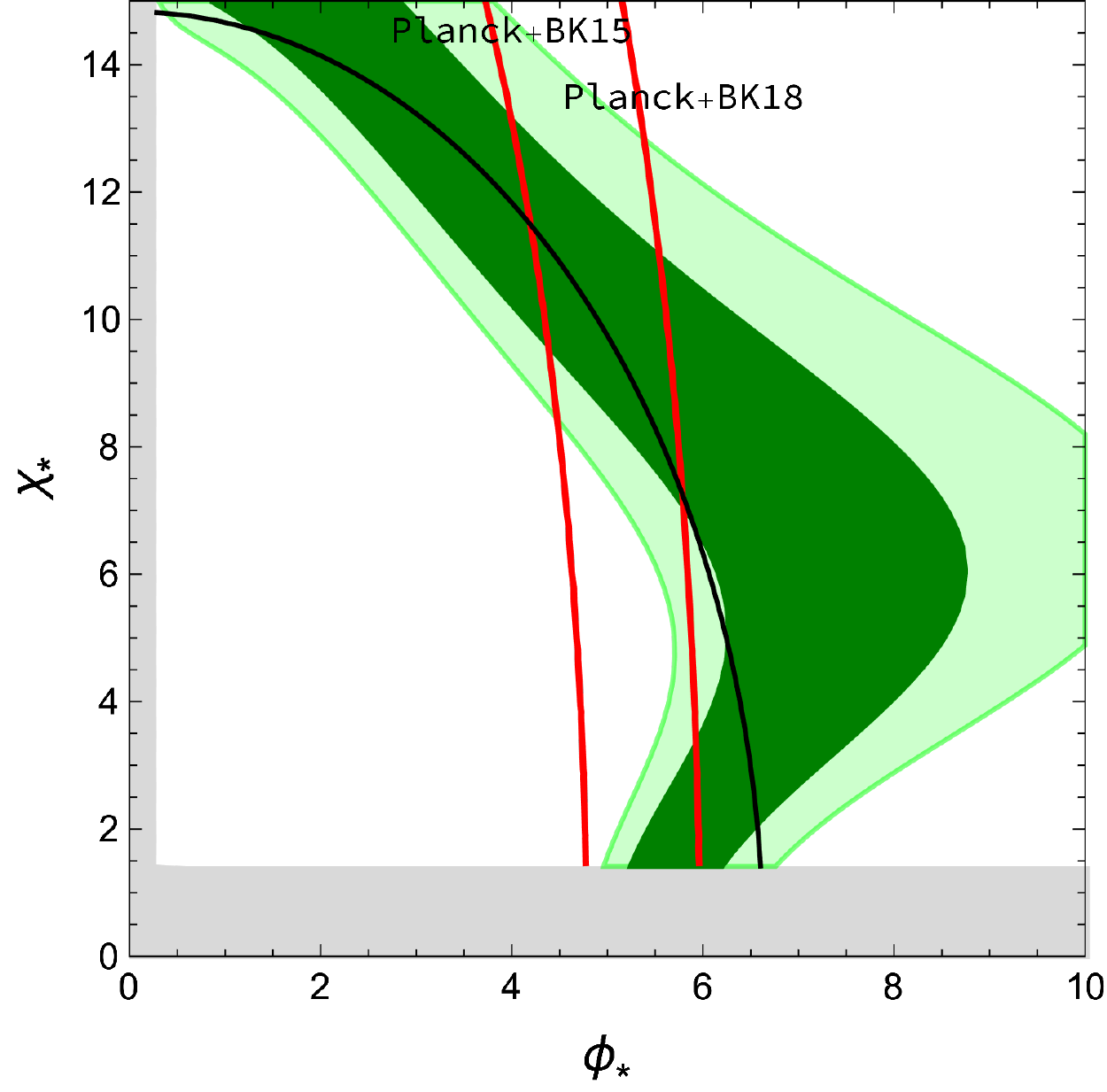}\end{minipage}
\end{tabular}
\caption{The left panel shows the prediction for $n_s$ and $r$ in the case of $p_\phi = 2/5$, $p_\chi = 2$, and $ \lambdaR = 1/20$.  
The black solid lines correspond to the $1\sigma$
and $2\sigma$ constraints from Planck+BICEP/Keck~2018~\cite{BICEP:2021xfz}.
The rainbow contour lines from left to right correspond to $N_*=50$ to 60. The color represents the change of the field value, $\phi_\ast$. The right panel is given by translating the result of $N_*=55$ in the left panel into the field value space. The black line lies on the sets of field values which realize $N_*=55$. 
The dark green and the light green region show the constraint for $n_s$ with $1\sigma$ and $2\sigma$ given by the Planck+BICEP/Keck~2018 result, which correspond to the values of $0.962<n_s<0.972$ and $0.958<n_s<0.976$, respectively. In the gray region, the slow-roll condition (\ref{double_baundary condition}) is broken. Right regions from two red lines correspond to the allowed range for $r$ given by the Planck 2018 (+BICEP/Keck 2015) result ($r<0.056$) and Planck+BICEP/Keck~2018 one ($r<0.036$).
}
\label{fig_diff_double}
\end{figure}
Up to here, we have considered the case that the power-law indices of the monomial potential for both fields are the same.
However, we can of course consider
different powers for two fields as
\begin{eqnarray}
V=U(\phi)+W(\chi)
=\lambda_\phi\frac{\phi^{p_\phi}}{\Mpl^{{p_\phi}-4}}+\lambda_\chi\frac{\chi^{p_\chi}}{\Mpl^{{p_\chi}-4}},\quad p_\phi \neq p_\chi.
\label{double_dif}
\end{eqnarray}
Let us briefly discuss the predictions in such a case.
For this potential, 
in the same way as in the case with the same power-law indices,
the spectral index $n_s$ and the tensor-to-scalar ratio $r$
can be approximately described as
\begin{eqnarray}
n_s - 1 &=& - \frac{1}{N_\ast} \left[ 1 + \frac{1}{N_\ast} \frac{\chi_\ast^2}{2 p_\chi \Mpl^2} \left( \frac{p_\phi}{p_\chi} -1 \right) \right]^{-1}
\frac{U(\phi_\ast) + \frac{p_\phi}{p_\chi} W(\chi_\ast)}{V(\phi_\ast,\chi_\ast)} - 2 \epsilon_\ast, \cr\cr
r &=& \frac{4 p_\phi}{N_\ast}  \left[ 1 + \frac{1}{N_\ast} \frac{\chi_\ast^2}{2 p_\chi \Mpl^2} \left( \frac{p_\phi}{p_\chi} -1 \right) \right]^{-1}.
\label{double_dif_nsr}
\end{eqnarray}
These expressions are actually reduced to Eqs.~\eqref{double_ns} and \eqref{double_r}
with $p_\phi = p_\chi = p$. 
From these expressions, we can find that in the case with $p_\phi \neq p_\chi$
the tensor-to-scalar ratio can move
between the values predicted by
the two single monomial models
with $p_\phi$ and $p_\chi$.
The spectral index is also
changed depending on the field trajectories.
As an example, in the left panel of Fig.~\ref{fig_diff_double},
we plot the predictions in the $n_s$-$r$ plane for the case with  $p_\phi = 2/5$, $p_\chi = 2$, and $\lambdaR = 1/20$. Here, the rainbow lines from left to right correspond to the cases with $N_\ast$ from $50$ to $60$.
The right panel shows the comparison of the predictions for $n_s$ and $r$ with the observational bounds in the $\phi_\ast$-$\chi_\ast$ field plane. The 1$\sigma$ and 2$\sigma$ constraints for $n_s$ are shown by the dark and light green regions, respectively. As for the tensor-to-scalar ratio, the red lines correspond to the upper bounds by the Planck+BICEP/Keck~2015 result ($r<0.05$) and Planck+BICEP/Keck 2018 one ($r<0.036$), and hence the right regions from the lines are allowed.
Since we fix $p_\phi, p_\chi$ and $\lambdaR$, the number of $e$-folds $N_\ast$ can be evaluated once we specify $\phi_\ast$ and $\chi_\ast$, which means that $n_s$ and $r$ become the function of  only field values by using Eq.~(\ref{double_dif_nsr}). Notice, however, that the field values $\phi_\ast$ and $\chi_\ast$ corresponding to $n_s$ and $r$ consistent with observations do not necessarily give $N_\ast = 50-60$. To check this, we also depict the combination of $\phi_\ast$ and $\chi_\ast$ giving $N_\ast=55$ in the right panel of Fig.~\ref{fig_diff_double} with black line. From the figure, one can see that the field values $\phi_\ast$ and $\chi_\ast$ (or the trajectories in the field space) should be carefully chosen in order to have $N_\ast=50$--$60$ and successful predictions for $n_s$ and $r$ simultaneously.
For the case considered here (i.e., $p_\phi=2/5$ and $p_\chi=2$), the potential energy is dominated by that of $\phi$-field
with $p_\phi = 2/5$ in the large region
in the field space, and hence
the predicted tensor-to-scalar ratio
is almost the same as for the case
with $p_\phi = 2/5$.
On the other hand, the spectral index
is relatively sensitive to the choice of the trajectories and its behavior is similar to that in the case with the same power-law indices as shown in Fig.~\ref{fig_double_nsr}.
Thus, from the viewpoint of constructing a model that is consistent with the current observational limit by minimal extension, even if we consider the double monomial inflation with the different power-law indices, the situation does not improve much from that with the same power-law indices.

\section{Double monomial inflation with spectator}
\label{triple}

As shown in the previous section, in the monomial case, that is, chaotic-type inflation, even if the two (or more) scalar fields can be introduced as the inflatons, the tensor-to-scalar ratio cannot become smaller than that in the single-field case with the same power-law index for the potential. This means that it is difficult to build a successful model with the scalar field having the potential $V(\phi) \propto \phi^p$~$(p\gtrsim 1)$ as the inflaton(s).

As a multi-field scenario, one can also consider the case where one of the scalar fields is the so-called spectator field which does not affect the inflationary dynamics, but can contribute to primordial density fluctuations such as in the curvaton scenario \cite{Lyth:2001nq, Enqvist:2001zp, Moroi:2001ct}, modulated reheating \cite{Kofman:2003nx, Dvali:2003em} and so on (see, e.g. \cite{Suyama:2010uj} for various models of this type). 
Indeed, in models with a spectator field, the tensor-to-scalar ratio can be suppressed compared to that in the single-field slow-roll inflation, which can make models with the potential $V(\phi) \propto \phi^p$~$(p\gtrsim 1)$ consistent with current observational bounds.
In the following, we assume that the spectator field $\sigma$ just has a mass term, that is, the potential is given by  $V(\sigma) = m_\sigma^2 \sigma^2/2$, where the mass  $m_\sigma$ is assumed to be
smaller than the Hubble parameter during the inflation. We have the curvaton scenario in mind, however, most arguments below hold for general spectator field ones.

The setup can be considered as a part of the multi-field model with separable-monomial potentials like the one discussed in the previous section in some way. To realize a curvaton scenario,
the main difference in the setup from the models discussed in the previous section
is that the system has a large hierarchy in the mass spectrum and the light scalar field is regarded as the curvaton. Furthermore, we assume that the curvaton  has 
a sub-Planckian field value\footnote{
When the curvaton has a super-Planckian field value, the second inflation can occur after the inflationary phase driven by the inflaton \cite{Vernizzi:2006ve, Moroi:2005kz, Dimopoulos:2011gb,Enqvist:2019jkb}. We do not consider such a case in this paper.
} (while other scalar fields which behave as the inflatons have super-Planckian field values).
In the curvaton scenario, during the inflation,
the curvaton slowly rolls by the Hubble friction
and can acquire the almost scale-invariant fluctuations originated from its quantum fluctuations.
After the end of the inflation, the Hubble parameter decreases, and
eventually, it becomes comparable to the mass of the curvaton, and then, the curvaton starts to oscillate around the potential minimum. The background dynamics of such an oscillating scalar field is similar to that of non-relativistic matter. During the radiation-dominated era after the completion of the reheating,
the ratio of the energy density of the oscillating curvaton to that 
of the radiation gradually increases. If the curvaton weakly couples to the radiation component, it can decay when the
Hubble parameter becomes almost comparable to its decay rate $\Gamma_\sigma$ and then the (isocurvature) fluctuations of the curvaton can be transferred into the adiabatic curvature perturbations. The amplitude of the generated curvature perturbations
is determined by the ratio of the energy density of the curvaton to the total at the decay.

Our following discussion on the $n_s$-$r$ plane applies not only to the curvaton $\sigma$ but also to a wide class of spectator field models. Thus $\sigma$ can be regarded as a general spectator field with a simple quadratic potential, which is characterized by its mass.

\subsection{Mixed-double model -- single-inflaton and single-spectator --}
\label{curvaton}

First, we review the predictions for the spectral index and the tensor-to-scalar ratio in the mixed-double model where a single inflaton and a single spectator can contribute
to the adiabatic curvature perturbations
(see, e.g., Refs.~\cite{Vernizzi:2006ve,Moroi:2005kz,Moroi:2005np,Ichikawa:2008iq,Ichikawa:2008ne,Enqvist:2013paa, Vennin:2015vfa,Fujita:2014iaa}). 
Here, following the setup in the previous section,
we simply consider the case where
the inflaton and the spectator have separable-monomial potential as 
\begin{eqnarray}
V=V(\phi)+V(\sigma)=\lambda\frac{\phi^p}{\Mpl^{p-4}}+\frac{1}{2}m_{\sigma}^2\sigma^2.
\label{curvaton_V}
\end{eqnarray}
As we have mentioned, the background dynamics of the inflation depends  only on the inflaton, so the number of $e$-folds and the slow-roll parameter $\epsilon$ are respectively written as Eqs.~(\ref{single_enumber}) and (\ref{eq:slowepsilon}). But in general, we have to consider the contributions to the curvature perturbations both from the inflaton and the spectator, which can be formally written by the sum of two field contributions as
\begin{eqnarray}
\zeta=\zeta^{(\phi)}
+
\zeta^{(\sigma)},
\label{curvaton_zeta}
\end{eqnarray}
and the power spectrum of the curvature perturbation is also given by a separable form as
\begin{eqnarray}
\mathcal{P}_{\zeta}&=&\mathcal{P}_{\zeta}^{(\phi)}+\mathcal{P}_{\zeta}^{(\sigma)}.
\label{curvaton_pzeta}
\end{eqnarray}
By introducing a parameter $R \equiv \mathcal{P}_{\zeta}^{(\sigma)}/\mathcal{P}_{\zeta}^{(\phi)}$
which represents the ratio of each contribution in the power spectrum, we get the formulae for $n_s$ and $r$ in the mixed-double model as~(see, e.g., Refs.~\cite{Vernizzi:2006ve,Moroi:2005kz,Moroi:2005np,Ichikawa:2008iq,Ichikawa:2008ne,Enqvist:2013paa,Fujita:2014iaa,Vennin:2015vfa})
\begin{eqnarray}
n_s-1
&=&-\frac{1}{\mathcal{P}_{\zeta}^{(\phi)}(1+R)}\left[\frac{d\mathcal{P}_{\zeta}^{(\phi)}}{dN_\ast}+\frac{d\mathcal{P}_{\zeta}^{(\sigma)}}{dN_\ast}\right] \nonumber \\
&=&\frac{1}{1+R}\left[-\frac{1}{N_\ast}-2\epsilon_\ast\right]
+\frac{R}{1+R}\left[-2\epsilon_\ast+\frac{2m_\sigma^2}{3H_*^2} \right] \nonumber \\
&=&-\frac{1}{1+R}\frac{2+p}{2N_
\ast}
+\frac{R}{1+R}\left[-\frac{p}{2N_\ast}+\frac{2m_\sigma^2}{3H_*^2}\right],
\label{curvaton_ns}\\
r&=&\frac{\mathcal{P}_T}{\mathcal{P}_{\zeta}^{(\phi)}(1+R)}
=\frac{16\epsilon_\ast}{1+R}
=\frac{16}{1+R}\frac{p}{4N_\ast},
\label{curvaton_r}
\end{eqnarray}
where we have used the expression \eqref{single_ns} for the spectral index of the power spectrum of the inflaton's contribution, $d \ln {\mathcal P}_\zeta^{(\phi)} / dN_\ast$.
Once the spectator model is specified, $R$ should be given in terms of the model parameters.
For example, in the standard curvaton scenario, $R$ is given by (e.g., Ref.~\cite{Enqvist:2013paa})
\begin{eqnarray}
&&R = \frac{8}{9} \epsilon_\ast f_\sigma^2 \left(\frac{\sigma_\ast}{\Mpl}\right)^{-2}, \cr\cr
&&f_\sigma \simeq \mathrm{Min} \left[ 1, \left( \frac{\sigma_\ast}{\Mpl} \right)^2 \left( \frac{m_\sigma}{\Gamma_\sigma} \right)^{1/2} \right],
\label{eq:fsigma}
\end{eqnarray}
where $\sigma_\ast$ is the field value of the curvaton at the horizon crossing during the inflation, and $f_\sigma$ roughly represents
the ratio of the energy density of the curvaton to the total one at the time of the curvaton decay.

As can be seen in the above expressions, the predictions are characterized by four parameters ($p, R, N_\ast, m_{\sigma}^2/H_\ast^2$). 
Note that the inflationary Hubble parameter, $H_*$, itself cannot be a free parameter and should be fixed to be consistent with the observed amplitude of the scalar curvature perturbations as
\begin{eqnarray}
H_*^2
=\frac{\pi^2 \Mpl^2}{2}\,r\times \mathcal{P}_{\zeta,{\rm obs}}
=\frac{\pi^2 \Mpl^2}{2}\frac{16}{1+R}\frac{p}{4N_\ast}\times \mathcal{P}_{\zeta,{\rm obs}} \,.
\label{curvaton_Hconstrain}
\end{eqnarray}
This just corresponds to fixing the size of $\lambda$ when $p, R$ and $N_\ast$ are given. From the Planck observation, the current mean value of the amplitude of the power spectrum is found to be ${\mathcal P}_{\zeta,{\rm obs}} = 2.2 \times 10^{-9}$~\cite{Planck:2018jri}.

\begin{figure}[t]
\begin{tabular}{c}
\begin{minipage}{0.45\hsize}\includegraphics[scale=0.5,clip]{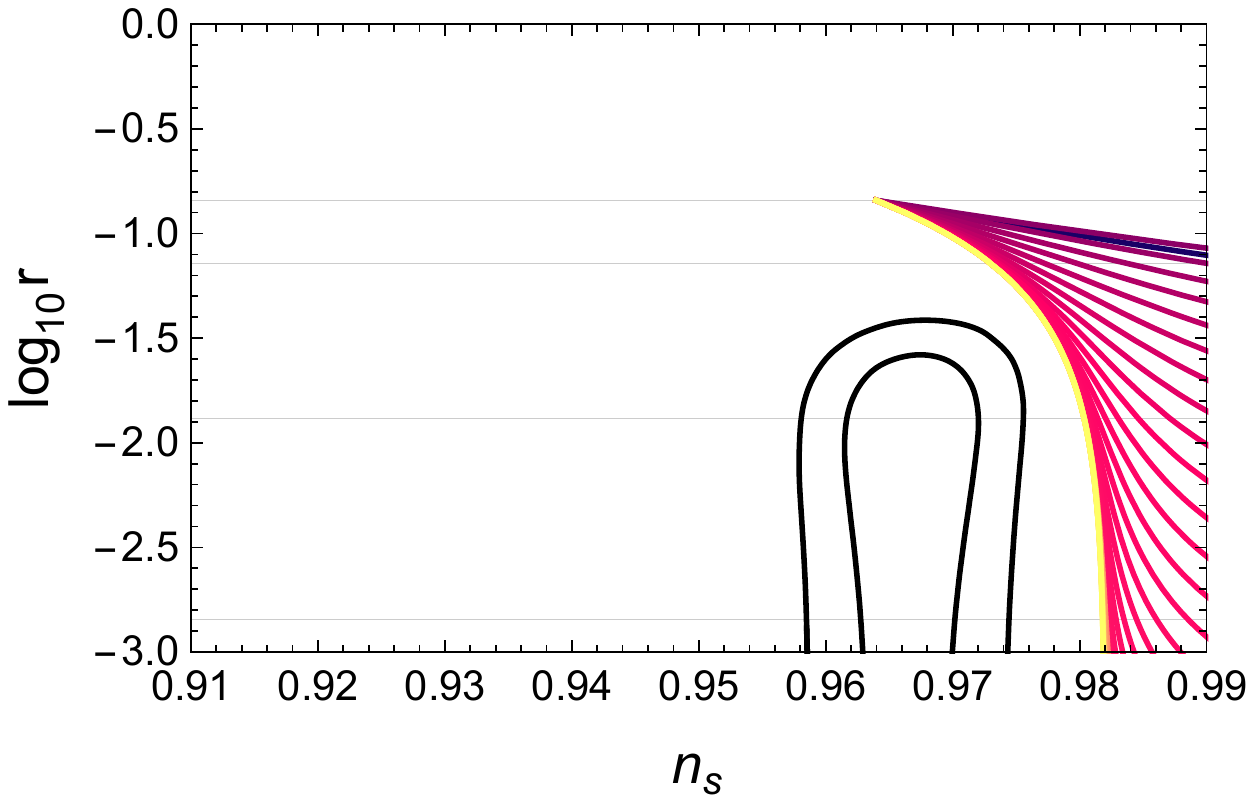}\end{minipage}
\begin{minipage}{0.45\hsize}\includegraphics[scale=0.5,clip]{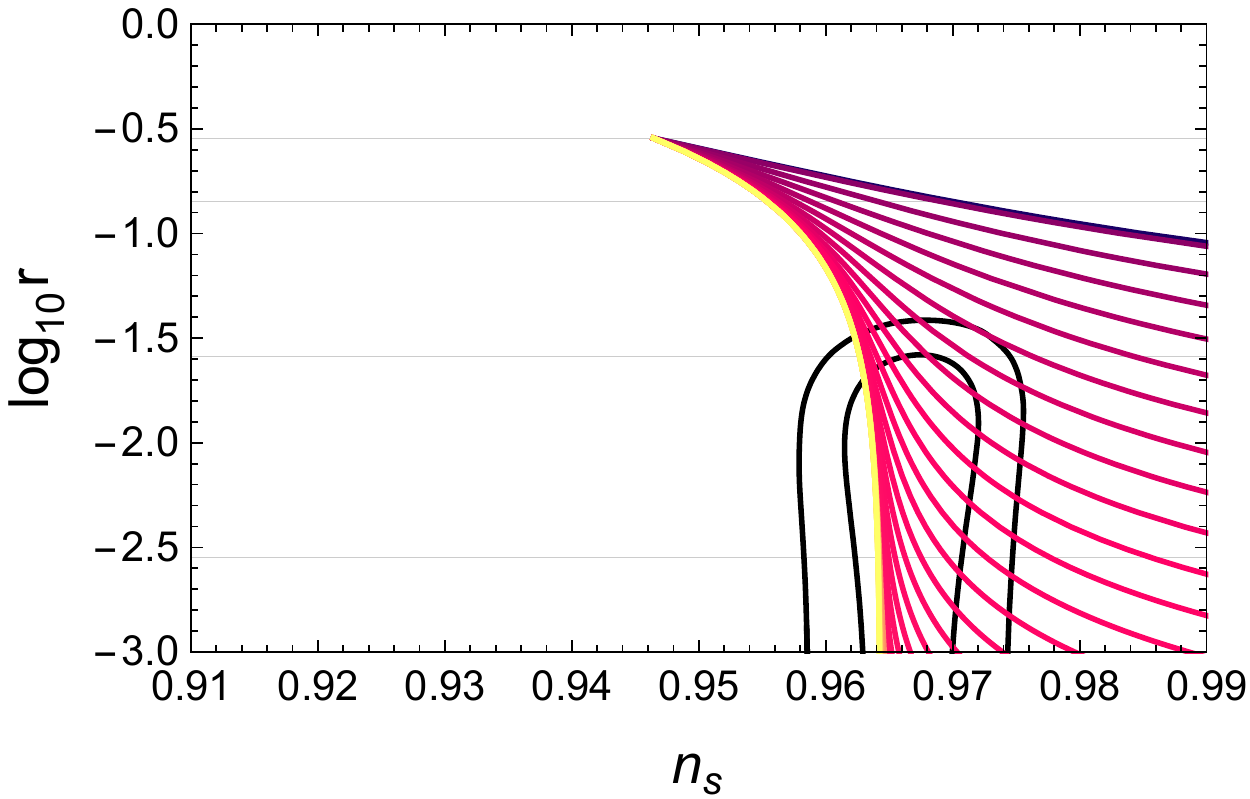}\end{minipage}
\begin{minipage}{0.1\hsize}\includegraphics[scale=0,clip]{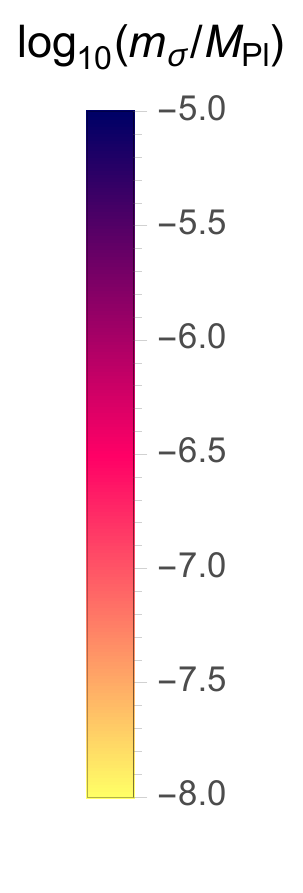}\end{minipage}\\
\begin{minipage}{0.45\hsize}\includegraphics[scale=0.5,clip]{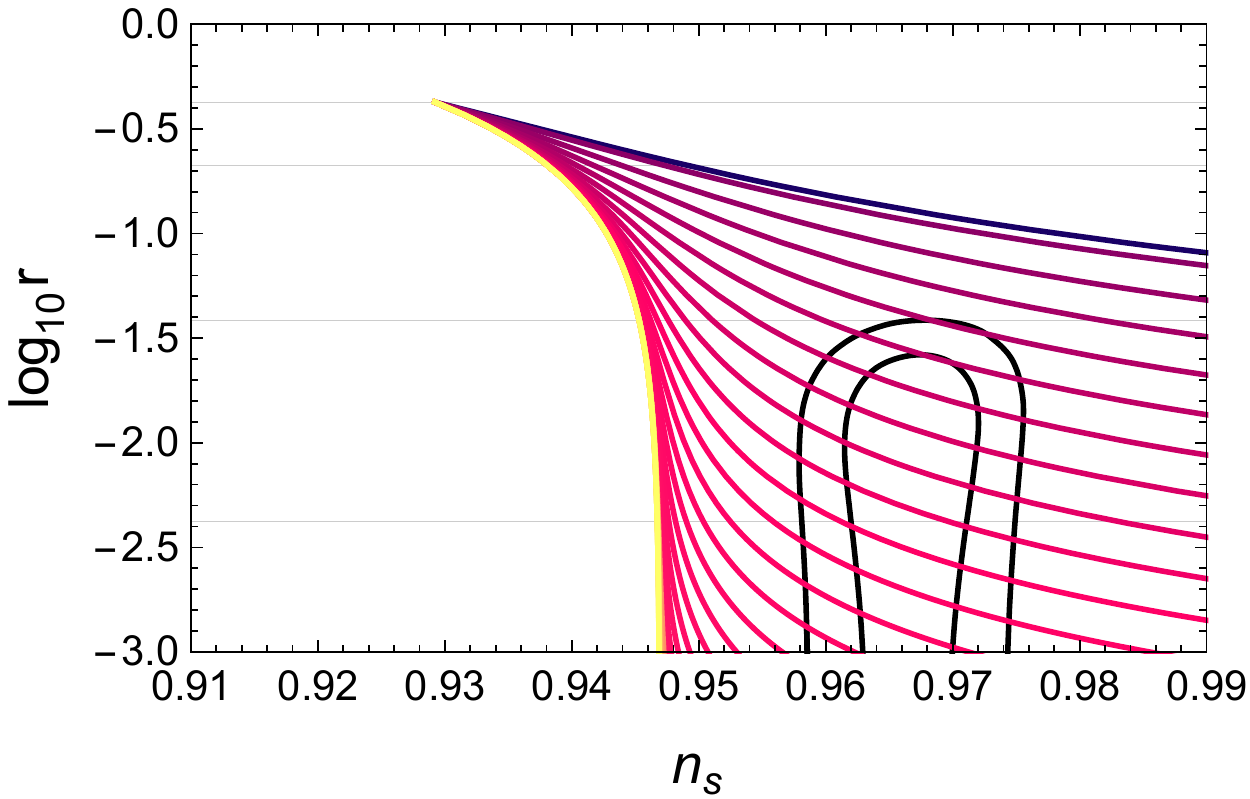}\end{minipage}
\begin{minipage}{0.45\hsize}\includegraphics[scale=0.5,clip]{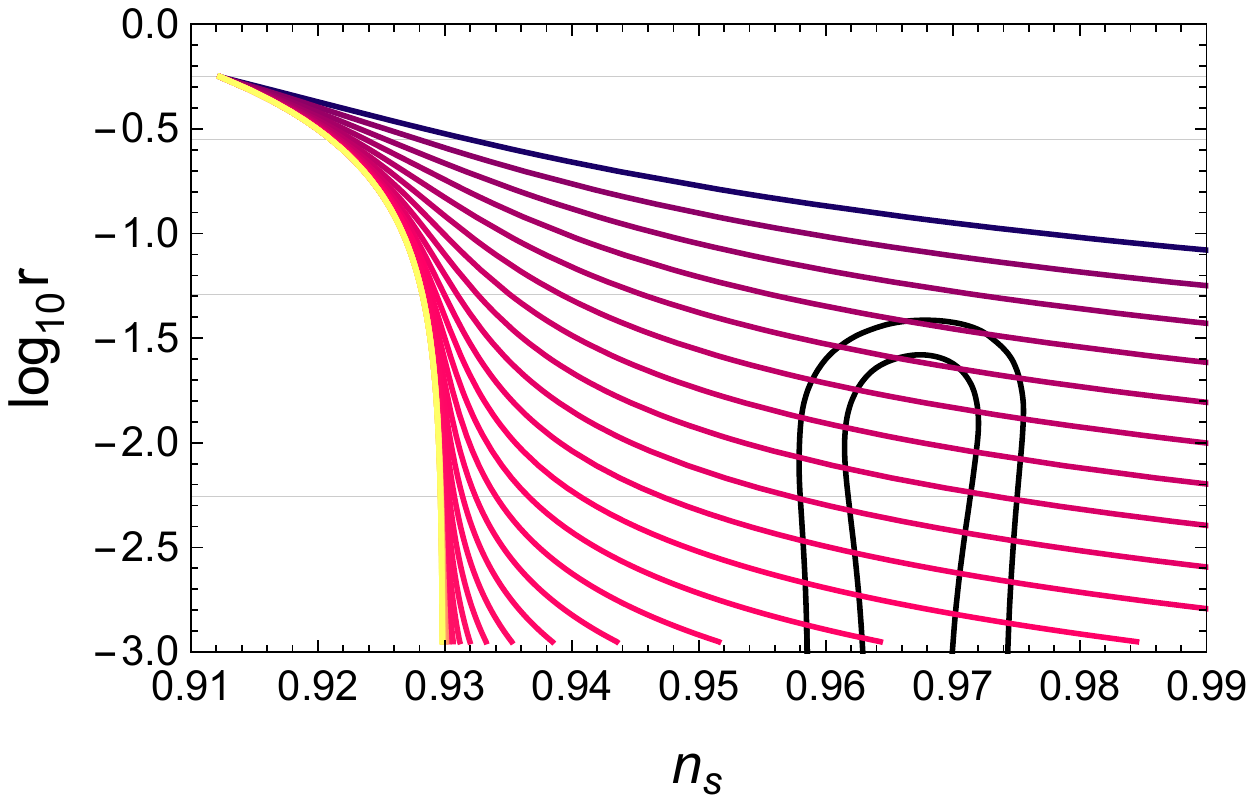}\end{minipage}
\begin{minipage}{0.1\hsize}\includegraphics[scale=0.5,clip]{figs/nsr_bar.pdf}\end{minipage}
\end{tabular}
\caption{Predictions for the mixed-double model in the $n_s$--$r$  plane. Cases with $p=2$ (upper left), $4$ (upper right), $6$ (lower left), and $8$ (lower right) are shown. The black contours indicate $1\sigma$ and $2\sigma$ allowed regions from the Planck+BICEP/Keck~2018 data \cite{BICEP:2021xfz}. The color lines correspond to the predicted value of $n_s$ and $r$, which are calculated from Eqs.~(\ref{curvaton_ns}) and (\ref{curvaton_r}), by varying $R$ as $R = 0-\mathcal{O}(10^2)$ from top to bottom, and  $m_\sigma$ as $m_\sigma /\Mpl = 10^{-5} \, - 10^{-8} \,$ whose values are indicated by color. The color bar legend indicates the value of  $\log_{10}(m_\sigma/\Mpl)$. The horizontal gray lines correspond to $R=0, 1, 10$ and $10^2$ from top to bottom which are depicted by using Eq. (\ref{curvaton_r}).}
\label{fig_curvaton_nsr}
\end{figure}

In Fig.~\ref{fig_curvaton_nsr}, by using the expressions (\ref{curvaton_ns}) and (\ref{curvaton_r}),
we plot the predictions in the $n_s$--$r$ plane
for the case with $p=2$ (upper left), $4$ (upper right), $6$ (lower left) and $8$ (lower right). The black contours correspond to $1\sigma$ and $2\sigma$ constraints from Planck+BICEP/Keck 2018 data \cite{BICEP:2021xfz}.  Here the number of $e$-folds is fixed to be $N_\ast = 55$, and then
in each panel we have one-to-one correspondence between the parameter $R$ and the tensor-to-scalar ratio $r$ which is nothing but the inflationary Hubble scale through the Eq.~\eqref{curvaton_Hconstrain}. In this figure, we also show the corresponding values of $R$ as horizontal gray lines. From top to bottom, we chose $R = 0$, $1$, $10$, and $10^2$.  Among them, the case of $R=0$ corresponds to the pure inflaton case where the contribution of the spectator to the curvature perturbations is negligible. The color lines from top to bottom are drawn by changing the mass
of the spectator field as $ m_\sigma /\Mpl = 10^{-5} 
- 10^{-8}\,$ in each panel.
From this figure, we find that we can rescue the models with $p=4,6$ and 8 by tuning $R$ and $m_\sigma^2$ appropriately. This is because the larger $R$ gives the smaller $r$ and positive contribution from $m_\sigma^2/H_\ast^2$ can make $n_s$ larger, i.e., shift to the right in $n_s$-$r$ plane.\footnote{
By introducing the Hubble induced mass based on the context of supergravity, it could be possible to realize the negative effective mass of the curvaton, and then it can make the spectral index shift to the left (see, e.g.,~\cite{Fujita:2014iaa}).
} However, for the case with $p=2$, the asymptotic value of $n_s$ in the limit of $m_\sigma^2 / H_\ast^2 \to 0$ is located outside the allowed region, and hence it is difficult to realize the consistent model with $p=2$ even if the contribution from the spectator field is included. Note that the case with $N_\ast=50$ also fails to be consistent with the observational result. 

\subsection{Hierarchical-triple model -- double-inflaton and single-spectator --}
\label{main}

As we have shown in Sec.~\ref{double_chaotic}, the double monomial inflation can decrease $n_s$ ($n_s$ can be shifted to the left in the $n_s$--$r$ plane) for the same monomial power. Thus the double-inflaton model with $p=2/5$ can be consistent with the recent Planck-BICEP/Keck result. However, $r$ in the double monomial inflaton model cannot be suppressed, and hence the case with $p=2$ cannot become viable. On the other hand, as shown in Sec.~\ref{curvaton}, the spectator model can decrease $r$, and hence the mixed-double model, that is, the single-inflaton and single-spectator model, with $p \ge 4$ ($p$ is the power-law index of the inflaton's potential) can become viable when the contribution from the spectator fluctuations is large enough although $n_s$  does not change much when the mass of the spectator field is negligibly small. Therefore, it is still difficult to rescue the simple quadratic inflationary model, $V(\phi) \propto \phi^2$ either in the double-inflaton model or the mixed-double one.
Here, in this section, we extend the above-mentioned models to
the three-field model with a hierarchical mass spectrum where two of them are the inflatons and one light scalar is the spectator field and show that the model with $V(\phi) \propto \phi^2$ can become consistent with current observational constraints in this framework.

Let us consider the massive three-field model whose potential is given by
\begin{eqnarray}
V=\frac{1}{2}m_{\phi}^2\phi^2+\frac{1}{2}m_{\chi}^2\chi^2+\frac{1}{2}m_{\sigma}^2\sigma^2.
\label{triple_V}
\end{eqnarray}
We assume that the inflationary phase proceeds by two inflatons, denoted by $\phi$ and $\chi$, and hence we can use Eq.~(\ref{double_epsilon2}) as the expression for the slow-roll parameter:
\begin{eqnarray}
\epsilon_\ast = \frac{1}{2N_\ast}\left[ 1 + \frac{\fR^2 \left( 1 - \mR^2 \right)^2}{ \left( 1 + \mR^2 \fR^2 \right)^2} \right],
\label{triple_epsilon}
\end{eqnarray}
where $N_\ast$ and $\fR$ are the ones introduced in Sec.~\ref{double_chaotic}, but now we fix $p=2$
and use $\mR=m_\chi/m_\phi$ instead of $\lambdaR$. 
As in the previous section~\ref{curvaton}, the power spectrum of the primordial curvature perturbation can be written by the sum of the contributions from three fields because of the separable potential as 
\begin{eqnarray}
\mathcal{P}_\zeta
&=&\mathcal{P}_\zeta^{(\phi)}+\mathcal{P}_\zeta^{(\chi)}+\mathcal{P}_\zeta^{(\sigma)}
=\mathcal{P}_\zeta^{\rm{(inf)}}+\mathcal{P}_\zeta^{\rm{(spec)}} \,,
\label{triple_pzeta}
\end{eqnarray}
where $\mathcal{P}_\zeta^{\rm{(inf)}} = \mathcal{P}_\zeta^{(\phi)}+\mathcal{P}_\zeta^{(\chi)}$ and $\mathcal{P}_\zeta^{\rm{(spec)}}= \mathcal{P}_\zeta^{(\sigma)}$. 
Then, following the discussions in the previous
sections \ref{double_chaotic} and \ref{curvaton},
$n_s$ and $r$ in this setup are respectively given by
\begin{eqnarray}
n_s-1
&=&\frac{1}{1+R}\frac{d\ln{\mathcal{P}_{\zeta}^{\rm{(inf)}}}}{d N_\ast}
+\frac{R}{1+R}\frac{d\ln{\mathcal{P}_{\zeta}^{\rm{(spec)}}}}{dN_\ast} \nonumber\\
&=&-\frac{1}{1+R}\left[2\epsilon_\ast+\frac{1}{N_\ast}\right]
+\frac{R}{1+R}\left[-2\epsilon_\ast +\frac{2m_\sigma^2}{3H_*^2} \right] \nonumber \\
&=&-2\epsilon_\ast -\frac{1}{1+R}\frac{1}{N_\ast}+\frac{R}{1+R}\frac{2m_\sigma^2}{3H_*^2}
\label{triple_ns},\\
r&=&\frac{\mathcal{P}_g}{\mathcal{P}_{\zeta}^{\rm{(inf)}}(1+R)}=\frac{1}{1+R}\frac{8}{N_\ast} \,.
\label{triple_r} 
\end{eqnarray}
where $R=\mathcal{P}_\zeta^{ \rm (spec)}/\mathcal{P}_\zeta^{\rm (inf)}$.  In the following calculation, we neglect
the contribution from the mass of the spectator field in $n_s$,
which is often assumed in a simple setup of spectator scenarios.

Now we need to mention the constraint on $\mathcal{P}_\zeta$. In the previous section, 
to normalize the amplitude of $\mathcal{P}_\zeta$ to the observed one ${\cal P}_{\zeta, {\rm obs}}$, we specified the inflationary Hubble
parameter via (\ref{curvaton_Hconstrain}) and considered $R$ as a free parameter.
But, here we fix the value of $R$ through the relation:
\begin{eqnarray}
1+R=\frac{\Mpl^2}{N_\ast}\left(\frac{2\pi}{H_*}\right)^2\times \mathcal{P}_\zeta,
\label{triple_Rfix} 
\end{eqnarray}
from which $R$ is determined by changing the Hubble parameter. The Hubble parameter at the horizon crossing, $H_\ast$, is determined by fixing the inflation parameters: the masses of the inflatons $(m_\phi, m_\chi)$
and also the values of the inflaton fields at the reference scale $(\phi_*, \chi_*)$, in other words, the choice of the background trajectory in the field space.
Note that through the expression of the number of $e$-folds
in the double inflation \eqref{double_enumber},  $(\phi_\ast, \chi_\ast)$ can be
written in terms of $(N_\ast, \fR)$. In the following analysis, 
we fix the number of $e$-folds at the horizon crossing to be $N_\ast = 55$ as a reference as in the previous section. To summarize, in this three-field models, we have four free parameters: $(m_\phi, m_\chi, \phi_\ast, \chi_\ast)$ or $(m_\phi, \mR, N_\ast, \fR)$. By fixing the number of $e$-folds, we can  reduce the parameters to $(m_\phi, \mR, \fR)$.

\begin{figure}[t]
\begin{tabular}{c}\includegraphics[scale=0.9]{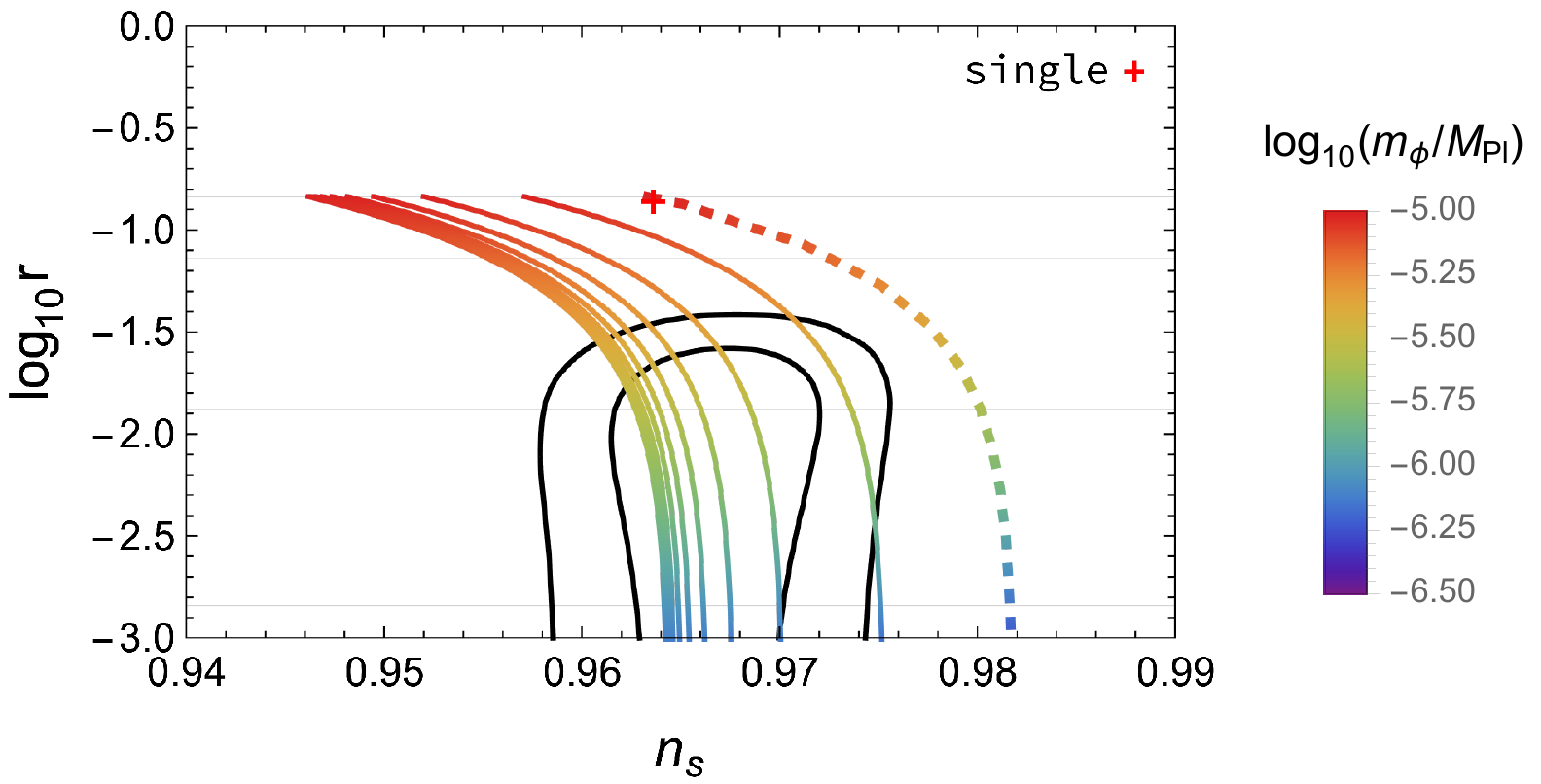}
\end{tabular}
\caption{$n_s$-$r$ plot for the case with $\fR = 1$. The change of $m_\phi$ is shown as the rainbow-color variation which is indicated by the bar legend for $\log_{10}(m_\phi/\Mpl)$. From right to left, each line corresponds to $\mR=1/n \, (n = 2,3, \cdots, 10)$. As a reference, the result for the standard single quadratic inflation is shown by the red cross point, and also that in the mixed-double model (the spectator is assumed to be almost massless) is shown by a dashed rainbow line extending from the red cross point, which is plotted by changing $R$. The rainbow-color variation of this dashed line is the same as the solid ones.
The gray horizontal lines indicate the value of $r$ which corresponds to $R=0,1,10$ and $10^2$ from top to bottom. The observational constraints from Planck+BICEP/Keck~2018 are shown as black lines as in Fig.~\ref{fig_curvaton_nsr}.
}
\label{fig_triple_nsr_mR}
\end{figure}

First, we investigate the $n_s$ and $r$ by changing $m_\phi /\Mpl =10^{-5}-10^{-6.5} $ and $m_R=1/2-1/10$ with $\fR = 1$. The result is shown in  Fig.~\ref{fig_triple_nsr_mR}.
In these cases, $m_\phi$ determines the energy scale of the inflation, that is, the Hubble parameter $H_\ast$, 
and then $R$ is determined through 
Eq.~\eqref{triple_Rfix}. Thus, in the figure, the smaller $m_\phi$ (whose actual values are shown by the rainbow color in the bar legend with respect to $\log_{10}(m_\phi/\Mpl)$) 
gives the larger $R$. For reference, we also show the gray horizontal lines for $R = 0,1,10$ and $10^2$ from top to bottom. Remind that a larger $R$ corresponds to a smaller tensor-to-scalar ratio. 
For the case of $f_R=1$, $\epsilon_\ast$ can be written, from Eq.~\eqref{triple_epsilon}, as
\begin{eqnarray}
\epsilon_\ast = \frac{1}{2N_\ast}\left[ 1 + \frac{\left( 1 - \mR^2 \right)^2}{ \left( 1 + \mR^2 \right)^2} \right],
\label{triple_epsilon_fR1}
\end{eqnarray}
from which one can see that the mass ratio $m_R$ in the inflaton sector is constrained by observational bound for a fixed $N_\ast$.
In the limit of $\mR \to 0$, $\epsilon_\ast$ for $\fR=1$ approaches to $1/N_\ast$, and then the spectral index is given by
\begin{eqnarray}
n_s - 1 \to  - \frac{3+2R}{1+R} \frac{1}{N_\ast}  ~.
\end{eqnarray}
Furthermore, by taking the limit of $R \gg 1$ in the above expression, we can find the attractor point for the spectral index $n_s$ as $n_s \to 1-2/N_\ast \simeq 0.964$ for $N_\ast = 55$. This prediction in the $n_s$--$r$ plane is similar to that in the so-called $\alpha$-attractor (see, e.g., Ref.~\cite{Kallosh:2013yoa}).
As one can see in Fig.~\ref{fig_triple_nsr_mR}, the hierarchical-triple quadratic model which consists of the two inflatons and the single spectator can be consistent with Planck observations
and the recent constraint on the tensor-to-scalar ratio obtained from BICEP/Keck 2018 data, $r_{0.05} < 0.036$ at $95\%$ C.L.~\cite{BICEP:2021xfz},
roughly indicates $m_\phi \lesssim 10^{-5.3} \Mpl$ which corresponds to $R \gtrsim 3$.

\begin{figure}[t]
\begin{tabular}{c}
\begin{minipage}{0.9\hsize}\includegraphics[scale=1]{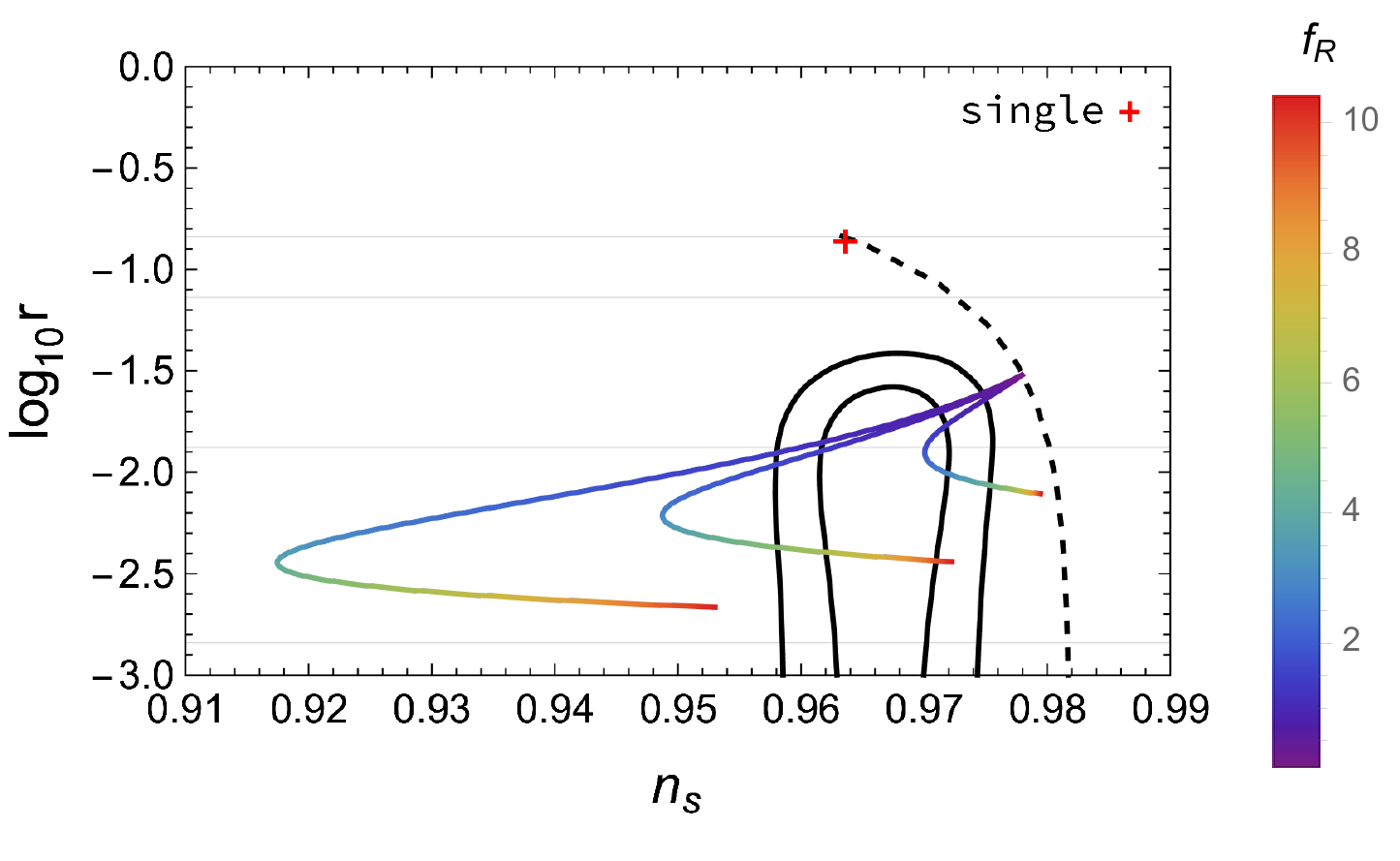}\end{minipage}
\end{tabular}
\caption{
$n_s$-$r$ plot for the hierarchical-triple quadratic model with the mass of the heavier inflaton as $m_\phi / \Mpl =3\times 10^{-6}$.
The rainbow color variation shows the change of $\fR$, and the correspondence between the color and the value of $\fR$ is shown in the bar legend. From right to left, each line is plotted for the cases with $m_R=1/2,\, 1/3$ and $1/4$. 
The meaning of the red cross, dashed line, gridlines, and black contours are the same as those in  Fig.~\ref{fig_triple_nsr_mR}.
}
\label{fig_triple_nsr_fR}
\end{figure}

Next, we discuss the dependence of the prediction in the $n_s$--$r$ plane on the choice of the background trajectory in the field space, that is, the value of $\fR (=\chi_\ast/\phi_\ast)$. Although in the above discussion, we have adopted $N_\ast$ and $f_R$ as model parameters instead of $\phi_\ast$ and $\chi_\ast$.  Then we also fixed them as $N_\ast=55$ and $\fR = 1$ as an example. However in fact, to realize $\fR = 1$ for $N_\ast = 55$,  we need to fine-tune the initial condition for the inflaton fields.
Therefore, to obtain a robust prediction in our model, it would be imperative to 
investigate how the result depends on such an initial condition by changing the value of $\fR$
and discuss the \textit{naturalness} of the prediction.
Here, as a reference,  we fix the mass of the heavier inflaton to be $m_\phi / \Mpl = 3\times 10^{-6}$ and $N_\ast = 55$,  which gives $r = O(10^{-2})$, and then consider the cases with $m_R = 1/2, 1/3$ and $1/4$. For each case, we vary the value of $\fR$ and investigate the predictions for $n_s$ and $r$. The result is shown in Fig.~\ref{fig_triple_nsr_fR}. Note that we have the limit for the range of the free parameter $f_R$ due to the slow-roll conditions for the two inflatons, $\phi$ and $\chi$, given by Eq.~(\ref{double_baundary condition}). 
As shown in this figure,
all cases can be consistent with the observation by choosing the value of $\fR$ properly. Notice that  $n_s$ has the minimum value depending on the value of $\mR$ because of the fact that the slow-roll parameter $\epsilon_\ast$ has the maximum at $\fR = 1/\mR$ as
\begin{eqnarray}
\epsilon_{\ast,\mathrm{max}} = \frac{1}{2N_\ast}
\frac{\left( 1 + \mR^2 \right)^2}{4 \mR^2}.
\end{eqnarray}
Thus, the smaller $\mR$ is, 
the larger the maximum value becomes, and hence $n_s$ can reach smaller values for smaller $\mR$ as seen in the figure. Therefore the range of $\fR$ which gives the observationally-consistent spectral index $n_s$ becomes smaller for the smaller $\mR$. 

\begin{figure}[t]
\begin{tabular}{c}
\begin{minipage}{0.3\hsize}\includegraphics[scale=0.45]{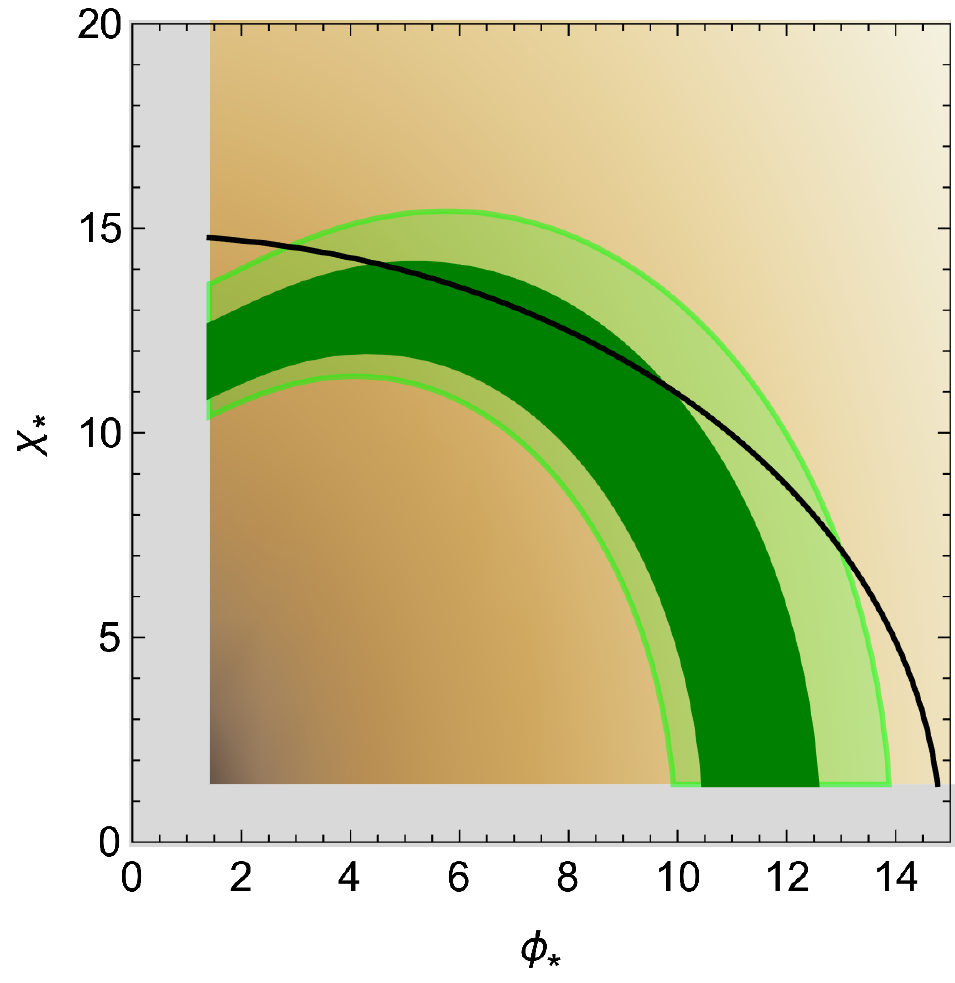}\end{minipage}
\begin{minipage}{0.3\hsize}\includegraphics[scale=0.45]{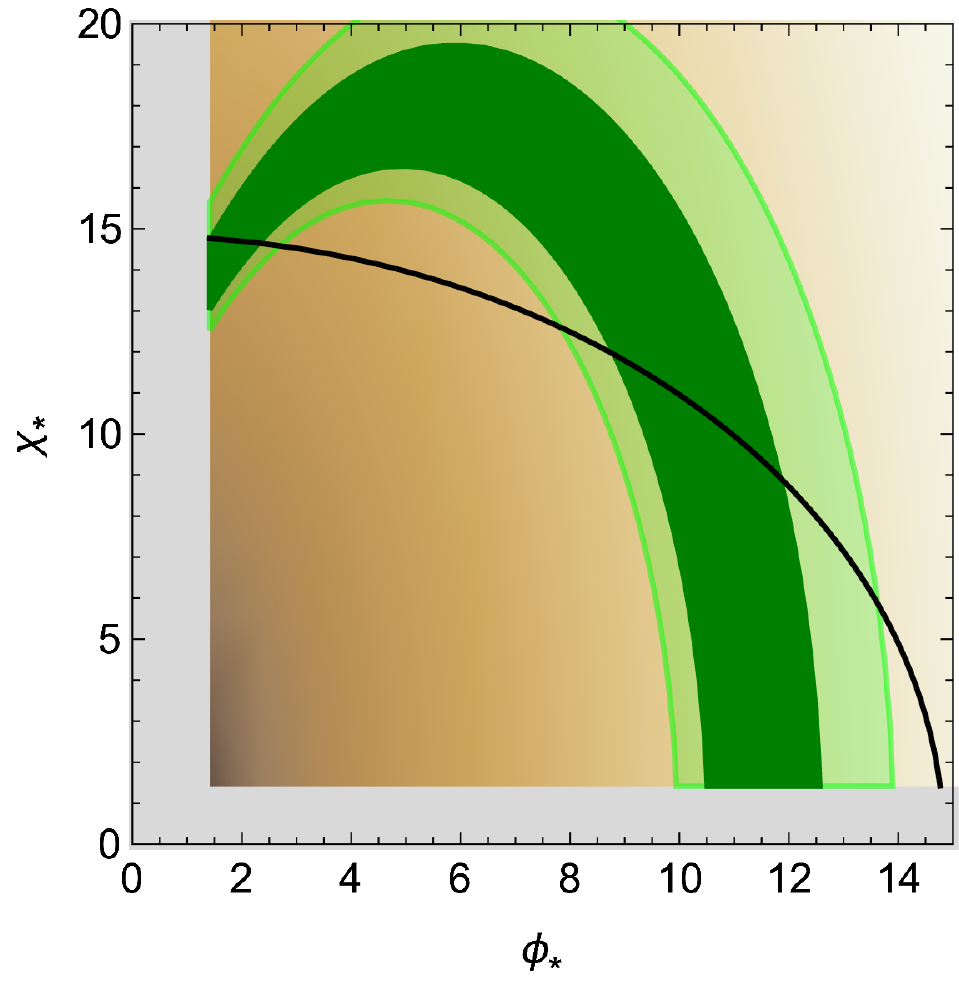}\end{minipage}
\begin{minipage}{0.3\hsize}\includegraphics[scale=0.45]{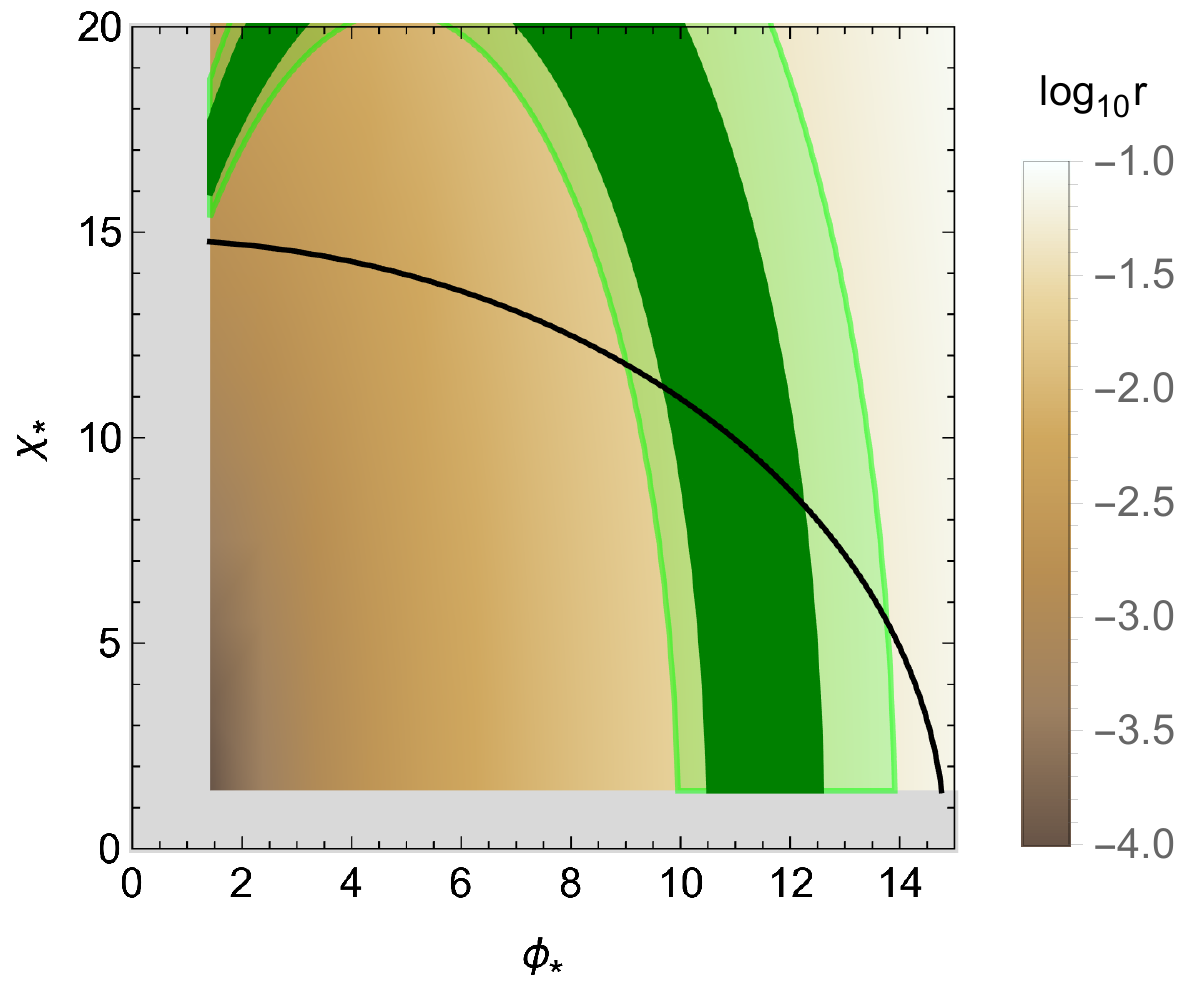}\end{minipage}
\end{tabular}
\caption{The values of $\phi_\ast$ and $\chi_\ast$ at the horizon exit corresponding to $N_\ast=55$ are depicted as the black line for the cases with $m_R=1/2$ (left), $1/3$ (middle) and $1/4$ (right). The bar legend on the right side denotes $\log_{10}r$. The dark and the light green regions show the $1\sigma$ and $2\sigma$ allowed ranges from Planck+BICEP/Keck~2018 data, which correspond to the values of $0.962<n_s<0.972$ and $0.958<n_s<0.976$, respectively. The region filling by gray is the region where the slow-roll condition is violated.}
\label{fig_triple_3nsr}
\end{figure}

In order to clarify the above argument, we plot the observationally-consistent region in the $\phi_\ast$- $\chi_\ast$ field plane in Fig.~\ref{fig_triple_3nsr}. As mentioned above, once $\phi_\ast$ and $\chi_\ast$ are fixed, $N_\ast$ and $f_R$ are determined, and hence $\phi_\ast$ and $\chi_\ast$ can be regarded as model parameters instead of $N_\ast$ and $\fR$. 
The dark (light) green region in each panel 
corresponds to that giving the observationally-consistent spectral index at $1\sigma$ ($2\sigma$) from Planck+BICEP/Keck~2018 data. On the other hand, the black line gives $N_\ast = 55$ via Eq.~\eqref{double_enumber}.
From this figure, we can find that, for the smaller $\mR$, the green region and the black line do not overlap much. This means that the smaller $\mR$ is, the more fine-tuning would be required for the choice of the initial conditions. Based on this result, we find that \textit{naturalness} favors the value of $\mR$ close to unity, i.e., the masses of the two inflatons may better be almost degenerate.

Although the above discussion is just for the case with $N_\ast=55$, we can roughly predict the result with other cases but around $N_\ast=55$ via Fig.~\ref{fig_triple_3nsr} since the change of $N_\ast$ can be described just by the shift of the black line in the figure.
Of course, the actual value of $n_s$ and $r$
should be dependent on the choice of $N_\ast$,
but the tendency should be the same.
Thus, we expect that even if $N_\ast = 50$ or $60$,  the favored values of $\mR$ would not change much.

As a conclusion for the three-field model, if the two scalars are inflatons and the other is the spectator field, the model can be consistent with the current observational constraint.
This requires the characteristic mass spectrum
where heavier two fields are almost degenerate and there is a large hierarchy between two heavy scalars and
the lightest scalar which can be regarded as
the spectator field.

\section{Discussion and conclusion}
\label{conclusion}

In this paper, we have investigated how additional scalar field(s) change the predictions for $n_s$ and $r$. As for the double-inflaton model, we have two choices for the assumption on the power-law index for the inflaton potential. For the first case, we choose the same power, and then $n_s$ can become smaller due to the existence of an additional term in the potential that contributes to the dynamics of the inflation, whereas $r$ takes the same value as the single-field model in the slow-roll approximation. As seen from Eq.~(\ref{double_epsilon2}), $\epsilon$ can only get larger than the single-field case. In particular, when $(p,\lambda_R)=(2,1)$ or $(f_R,\lambda_R)=(1,1)$, $\epsilon$ remains the same as the single-field result.
Therefore, the double monomial model can decrease the value of $n_s$ and,
for the case of $p=2/5$, $n_s$ can reach the Planck-BICEP/Keck~2018 allowed region in some range of $N_\ast$. In the second case, we choose different power-law indices for each inflaton potential. In this setup, we can get a consistent prediction with the observational constraint depending on the choice of the trajectory. However, to realize such a prediction, we need large hierarchical constant (coupling) parameters between two inflatons, and the total energy density is almost dominated by an inflaton with the lower power-law index. This means that, even if we consider a different power-law index for two inflatons, the setup is almost the same as the single-field case when the model becomes consistent with the observational constraint. Hence this extension does not improve the model much and is not so attractive compared to the same power-law case.
As the other two-field model, we also briefly reviewed the single-inflaton and spectator model, called mixed-double one, in which
the field value of the spectator is supposed to be smaller than the inflaton one, and its mass is assumed to be much lighter. Because of this, the contribution to the energy density is negligible during the inflation and it does not affect the inflationary dynamics. However, its contribution to the curvature perturbation can become important. 
Indeed, when the spectator's contribution to $\mathcal{P}_\zeta$ is large, it is known that we can decrease $r$ to make the model consistent with the observational bound for some monomial inflaton potential case. 

Based on these features in the extension of the multi-scalar model, we carefully investigated
the three-field model with double inflatons and single spectator in light of the recent Planck-BICEP/Keck 2018 result. This model contains the characteristics of the above two models, and hence $n_s$ can get decreased and $r$ can be suppressed. Indeed by appropriately choosing the model parameters in this three-field framework, even quadratic chaotic inflation can be rescued, which was impossible in the double-inflaton and mixed-double models.  As discussed in Sec~\ref{main}, to obtain successful predictions for $n_s$ and $r$, it is preferable to have almost degenerate masses for the two inflatons to avoid much fine-tuning of the trajectory of the fields. Therefore, three scalar fields with the quadratic potentials for each in this setup should have the mass hierarchy such that two heavy fields have almost degenerate masses and the other one has a much lighter mass. 
In fact, Ref.~\cite{Ellis:2013iea}
has also discussed the possibility of
rescuing the quadratic inflation in the three-field framework. By employing the Monte Carlo method, the predictions for
$n_s$ and $r$ were explored, but the dependence of the inflatons' parameters, in particular, the choice of the trajectory in the field space, which are focused in our work, on the observables has not been much discussed.

Our three-field model may also predict 
relatively large local-type primordial non-Gaussianities 
as in 
other multi-field models. While, as shown in Ref.~\cite{Yokoyama:2007dw}, the non-Gaussianity generated during inflation
is suppressed by the slow-roll parameters even in the multi-inflaton case,
the spectator field can be a source of large local-type non-Gaussianities.  
Thus the non-linearity parameter such as $f_{\rm NL}$ can be estimated as in the mixed inflaton and spectator models~(see, e.g., Ref.~\cite{Ichikawa:2008iq,Ichikawa:2008ne,Suyama:2010uj,Enqvist:2013paa}). 
For example, when considering the curvaton mechanism as a spectator field model, 
$f_{\rm NL}$ can be 
given by
\begin{eqnarray}
f_{\rm NL} \approx \frac{5}{6} \left( \frac{R}{1+R}  \right)^2 \left[ \frac{3}{2 f_\sigma} - 2 - f_\sigma \right]\,.
\end{eqnarray}
In our model, as discussed in section \ref{main},
the BICEP/Keck 2018 result, $r_{0.05} < 0.036$ at $95\%$ C.L.,
implies $R \gtrsim 3$ for $N_\ast \simeq 50 - 60$.
Taking this constraint into account, the Planck result for the primordial non-Gaussianity, $f_{\rm NL} = -0.9 \pm 5.1$ at $68\%$ C.L., gives $f_\sigma \gtrsim 0.1$.
These constraints on $R$ and $f_\sigma$ still be compatible with the discussion in section \ref{main}.

In fact, other extensions of the multi-scalar model have been proposed to realize the successful inflationary model consistent with the Planck observations. For example,
Refs.~\cite{Renaux-Petel:2014htw,Achucarro:2019pux} proposed 
models with 
the non-canonical kinetic term.
It would be interesting to investigate the possibility of distinguishing between our model and other models by scrutinizing the non-Gaussianity or fine features in the power spectrum.

Finally, we comment on the implication of our study to particle physics models, in particular, related to sneutrino (for possible connection, see e.g., \cite{Murayama:1992ua,Murayama:1993xu,Kawasaki:2000ws,Ellis:2003sq,Ellis:2004hy,Kadota:2005mt,Pallis:2011ps,Mazumdar:2010sa,Yamaguchi:2011kg,Haba:2011uz,Khalil:2011kd,Senoguz:2012iz,Nakayama:2013nya,Ellis:2014rxa,Murayama:2014saa,Harigaya:2014bsa,Ellis:2014xda,Nakayama:2016gvg,Kallosh:2016sej,Bjorkeroth:2016qsk,Gonzalo:2016gey,Nakayama:2017cij,Garg:2017tds,Haba:2017fbi}). In a supersymmetric model, there exists the so-called ``sneutrino", as scalar partner of right-handed neutrinos. There are naturally three sneutrinos and their potentials are quadratic ones, and hence three scalar fields in our three-field model may be identified with sneutrinos. As we have discussed, the masses of three scalar fields should have some hierarchy, which would give interesting implications to the model building of sneutrino physics.
Among the literature, e.g., Ref.~\cite{Haba:2017fbi} has investigated 
mixed single-inflaton and curvaton scenario in the sneutrinos framework. We leave the extension of this previous work to our three-field model as a future issue.

\section*{Acknowledgement}
\label{ack}

We would like to thank Yuichiro Tada and Takeshi Kobayashi
for useful discussion.
This work was supported in part by JSPS KAKENHI Grant Nos.~JP20H01932 (S.Y.), JP20K03968 (S.Y.), 17H01131 (T.T.) and  19K03874 (T.T.), and MEXT KAKENHI Grant No.~19H05110 (T.T.).

\bibliographystyle{apsrev4-1}
\bibliography{bib}

\end{document}